\documentclass[journal]{IEEEtran}

\usepackage{algorithm}
\usepackage{algorithmic}
\usepackage{amsmath,amssymb,amsthm,amsfonts}

\usepackage{graphicx}
\usepackage{epstopdf}
\usepackage{caption}
\usepackage{subcaption}
\usepackage{times}
\usepackage{multirow}
\usepackage{booktabs}
\usepackage{lscape}
\usepackage{color}

\graphicspath{ {./figures/}}

\ifCLASSINFOpdf
\else
\fi

\begin{document}

\title{Adapting Stochastic Block Models \\ to Power-Law Degree Distributions}

\author{Maoying~Qiao, Jun~Yu,~\IEEEmembership{Member,~IEEE,} ~Wei~Bian,~\IEEEmembership{Member,~IEEE,}
~Qiang~Li,~and~Dacheng~Tao,~\IEEEmembership{Fellow,~IEEE}% <-this % stops a space
\thanks{M. Qiao and J. Yu are with the School of Computer Science and Technology at Hangzhou Dianzi University, Hangzhou, China. (email: qiaomy@hdu.edu.cn, yujun@hdu.edu.cn). }
\thanks{ W. Bian is with the Centre for Artificial Intelligence in the Faculty of Engineering and Information Technology at the University of Technology Sydney (email: brian.weibian@gmail.com).}
\thanks{ Q. Li is with the School of Software in the Faculty of Engineering and Information Technology at the University of Technology Sydney, and the Department of Computing at Hong Kong Polytechnic University (email: leetsiang.cloud@gmail.com).}
\thanks{D. Tao is with the UBTECH Sydney Artificial Intelligence Centre and the School of Information Technologies in the Faculty of Engineering and Information Technologies at University of Sydney, 6 Cleveland St, Darlington, NSW 2008, Australia (email: dacheng.tao@sydney.edu.au).}
%\thanks{\textit{Corresponding author: Jun Yu.} }
%\thanks{Manuscript received **, 20**; revised **, 20**.}
\thanks{\textcopyright 2019 IEEE. Personal use of this material is permitted. Permission from IEEE must be obtained for all other users, in any current or future media, including reprinting/republishing this material for advertising or promotional purposes, creating new collective works, for resale or redistribution to servers or lists, or reuse of any copyrighted component of this work in other works.}
}

\markboth{IEEE TRANSACTION ON CYBERNETICS}%
{Shell \MakeLowercase{\textit{et al.}}: Bare Demo of IEEEtran.cls for IEEE Journals}

\maketitle

\begin{abstract}
Stochastic block models (SBMs) have been playing an important role in modeling clusters or community structures of network data. But, it is incapable of handling several complex features ubiquitously exhibited in real-world networks, one of which is the power-law degree characteristic. To this end, we propose a new variant of SBM, termed power-law degree SBM (PLD-SBM), by introducing degree decay variables to explicitly encode the varying degree distribution over all nodes. With an exponential prior, it is proved that PLD-SBM approximately preserves the scale-free feature in real networks. In addition, from the inference of variational E-Step, PLD-SBM is indeed to correct the bias inherited in SBM with the introduced degree decay factors. Furthermore, experiments conducted on both synthetic networks and two real-world datasets including Adolescent Health Data and the political blogs network verify the effectiveness of the proposed model in terms of cluster prediction accuracies.
\end{abstract}

\begin{IEEEkeywords}
Stochastic block models, power-Law degree distribution, EM algorithm.
\end{IEEEkeywords}

\IEEEpeerreviewmaketitle

\section{Introduction}
\label{sec:introduction}
Recent years, network data has been growing rapidly in a wide range of areas, from  bioinformatics to academic collaboration. How to mining useful information from such data, i.e., network analysis, has attracted both theoretical and computational studies   \cite{Airoldi2007StatNetAna}\cite{goldenberg2010survey}\cite{sarkar2011theoretical}\cite{ZafaraniUser2015}. Statistical models, with statistical inference, deal with uncertainties elegantly and thus provide powerful tools for network analysis such as disclosing hidden structures within the data. Although classical statistical models for network modeling have been explored for decades, e.g., exponential random graph model \cite{goldenberg2010survey}, latent space model \cite{hoff2002latent} and stochastic block model (SBM) \cite{snijders1997estimation}, studies show new features of real-world networks that cannot be fully modeled with existing approaches. Such ubiquitous features include small world phenomenon, power-law degree distributions, and overlapped cluster or community structures \cite{Prat-PerezPrat2016}\cite{changcommunity}\cite{yang2015unified}. Dozens of recent works have focused on modifying classical statistical models to improve the model capability and consequently improve the performance of statistical inference.

Stochastic block models (SBM) has been a significant statistical tool for latent cluster discovering in network data \cite{snijders1997estimation}\cite{holland1983stochastic}\cite{wang1987stochastic}, and a variety of its extensions have also been developed. On the assumption that the nodes of a network is partitioned into different clusters and the existence of edges between pairwise nodes depends only on the clusters they belong to,  Snijders and Nowicki \cite{snijders1997estimation} first proposed using posterior inference to uncover such cluster structures. Incorporating nonparametric Bayesian techniques in SBM, with a Chinese Restaurant Process (CRP) prior over the node partition imposed, \cite{kemp2006learning} addresses the issue of cluster number selection. In addition, mixed membership SBM (MMSBM) has also been developed \cite{airoldi2008mixed} to deal with the situation where nodes can have multi-label properties, namely belonging to overlapped clusters.

Other main extension of SBM includes hierarchical SBM \cite{Ho2012_MultiScaleSBMs}, integrating node attributes into SBM \cite{kim2012nonparametric}, dynamic infinite extension of MMSBM \cite{fan2013dynamic}, and improving model scalability by stochastic variational methods \cite{Chan:2011:ISF:2283516.2283600}\cite{peng2015scalable}. Due to its computational flexibility and structural interpretation, SBM and its extension have been popularizing in a variety of network analysis tasks, e.g., uncovering social groups from relationship data \cite{nowicki2001estimation}\cite{Wei:2015:MTP:2808688.2749465}\cite{Zhou:2015:SIB:2808688.2717314}\cite{liu2014multiobjective}, functional annotation of protein-protein interaction networks \cite{airoldi2008mixed}, and network clustering \cite{VuHunter2012}.

It has been long noticed that real networks exhibit a ubiquitous scale-free property, i.e., the distribution of node degrees following a power-law \cite{barabasi1999emergence}. For example, some nodes in the World Wide Web have far more connections than others and are recognized as ``hubs''. However, the traditional SBM is incapable to handle this naturally existing scale-free property in networks. This is due to its block property, which tends to grouping nodes of similar degrees rather than nodes whose degrees distribution follows a power-law. Thus, it can cause significant bias when applied to infer cluster structures especially for assortative networks. In this paper, a new extension of SBM is proposed, termed power-law degree stochastic block model (PLD-SBM), to address this problem. PLD-SBM explicitly encodes the power-law characteristic, and thus is capable to correct the bias caused by statistical inference in the traditional SBM. Its learning and inference are derived with efficient variational methods. We evaluate the proposed PLD-SBM on both simulated networks and two real networks.

The rest of the paper is organized as follows:
Section \ref{sec:relatedWork} reviews related literature.
Section \ref{sec:model} introduces our proposed model and analyzes the degree characteristic.
Section \ref{sec:algorithm} describes the learning and inference algorithms in PLD-SBM.
We present the experimental results of simulations in Section \ref{sec:simulations}, followed by real-world experiments on the Adolescent Health Network and the political blog network  in Section \ref{sec:adolescent} and \ref{sec:other}.
Section \ref{sec:conclusion} concludes this paper.

\section{Related Work} \label{sec:relatedWork}

A stochastic block model (SBM) handles the task of recovering blocks, groups or community structures in networks \cite{shiga2012variational}\cite{bu2016local}.

From a generative perspective \cite{yang2015emphasizing}, SBM, with given parameters specifying cluster-level node portions and edge connectivity, produces a network via two steps: firstly nodes of a pair are independently sampled from a multinomial prior over node portions, and secondly the existence of the edge between them is sampled from a prior distribution over connectivity. Different priors have been explored for certain situations, such as Bernoulli for simple graphs, a Poisson distribution for multigraphs, \cite{yan2014model}\cite{tang2012identifying}, and exponential family for valued graphs \cite{mariadassou2010uncovering}. Literally, SBM models the cluster structures in a block level and does not take the individuality of nodes into consideration. In other words, it treats nodes within a group equally. However such a strategy renders SBM incapable of handling node degree heterogeneity. For example, in the political blog network \cite{adamic2005political}, within either the liberal community or the conservative one, both popular blogs and inactive ones naturally exist, and the popular ones typically have higher degree than those of inactive. In this situation, the node degrees are heterogeneously distributed. However, according to the experimental results in \cite{karrer2011stochastic}, SBM fails to discover such clusters. Another similar example is in the clustering of Zachary karate network  \cite{leger2014detection}. A community in the club is naturally constituted with high-degree leading members such as instructors and low-degree satellite members such as students. The experiments show that SBM divided club members into degree-homogeneous groups, even though it can indeed indirectly identify degree-heterogeneous groups with extra steps by first finely dividing the network (i.e., with large group number) and then merging the high-degree instructor nodes and low-degree student nodes into groups of degree heterogeneity.

A degree-corrected SBM (DC-SBM) model to alleviate the heterogeneous bias inherited by SBM was proposed by Karren and Newman \cite{karrer2011stochastic}. An expected degree parameter is introduced for each node into a Poisson-valued SBM in order to handle the heterogeneity of node degrees.
Then, it smoothly incorporates the newly introduced node-wise degree parameters into the existing block parameters as new expected number of links via multiplying them together. Still, the derived objective function is only dependent on group-level degrees, rather than degrees of node-level.
Finally, an iterative process involving node switching groups is derived to attain a best cluster structure. On one hand, DC-SBM is superior to SBM in terms of capacity to handle degree heterogeneity within each cluster. Specifically, SBM tends to discover homogeneous structures while DC-SBM is able to split networks into heterogeneous groups via adding independent degree-encoded variables to each node. On the other hand, as mentioned above, we notice that its inference procedure is rather directly dependent on the introduced node-level expected degree parameters. Thus, there are no ``posterior'' explanations for the proposed model, and no network generation procedure can be derived. Besides, obtaining an optimal community structure is inefficient due to the almost exhaustive iterative procedure. Note that several studies have extended DC-SBM by considering different modeling scenarios, such as model selection \cite{yan2014model}, clustering consistency \cite{zhao2012consistency}, edge direction type \cite{zhu2014oriented} and regularized spectral analysis \cite{qin2013regularized}.

Latent degree learning has also been investigated to address the heterogeneous bias of SBM. One typical approach is a Link Density model (LD) \cite{morup2009learning} built on MMSBM \cite{airoldi2008mixed}. Essentially, MMSBM contains two layers of variables - one observation layer representing edges and one latent layer denoting nodes. To model degree heterogeneity, LD adds another latent layer, representing link properties, into the standard MMSBM. The probability of each latent link variable is decomposed into a product of two free parameters over node-specific degrees.
Finally, model parameter learning and MAP inference are solved via a variational Bayesian approach.
Though at first glance our proposed model seems to be similar to LD, there exist fundamental differences, namely node-specific degree parameters vs node-specific latent degree variables.
The latter one is able to incorporate a power-law prior distribution into SBM to handle the scale free characteristic, which is ignored by the former LD model.

There is also other literature on addressing degree heterogeneity in networks using spectral methods, for which we leave the references \cite{reichardt2011interplay}\cite{chaudhuri2012spectral}\cite{dasgupta2004spectral}\cite{tao2007general} for interested readers.

\section{The Proposed Model}
\label{sec:model}

SBM constructs networks by two matrices and a layer of hidden variables. One matrix is an $N\times N$ binary adjacency matrix $Y$, representing observed edge connectivity of an undirected binary random graph $\mathcal G$. Its entries $\{y_{ij}\}$ with value $1$ or $0$ indicates the existence or nonexistence of an edge between two nodes $i$ and $j$. The other matrix is a $K\times K$ matrix $B$, parameterizing the connection probabilities between clusters, i.e., a node from cluster $k$ is connected to a node from cluster $k'$ with probability $0\le b_{kk'}\le 1$. In addition, the cluster or community structure underlying a network is encoded via an extra layer of hidden variables, each of which $z_i\in\{1,2,...,K\}$ associated with one node and whose value within $K$ indicating its belonging of clusters or communities. Given nodes' cluster indexes, the single-edge likelihood for each observed edge $y_{ij}$ is given below, and the whole-edge likelihood is easily obtained by multiplying over the whole edge set.
  \begin{align}\label{eq:SBMs}
     p(y_{ij}=1|z_i,z_j,B)=b_{z_iz_j}.
  \end{align}
A multinomial prior $\mbox{Multi}(\pi)$ is assigned to each latent variable $z_i$. Maximum likelihood estimation (MLE) or Bayesian methods can be applied to learn the model parameter $B$, and then a posterior inference on $\mathbf z$ reveals the network's cluster structure, i.e., the cluster indexes of nodes in the network \cite{snijders1997estimation}.

\begin{figure*}[thb]
                \centering
                \includegraphics[width=0.7\textwidth]{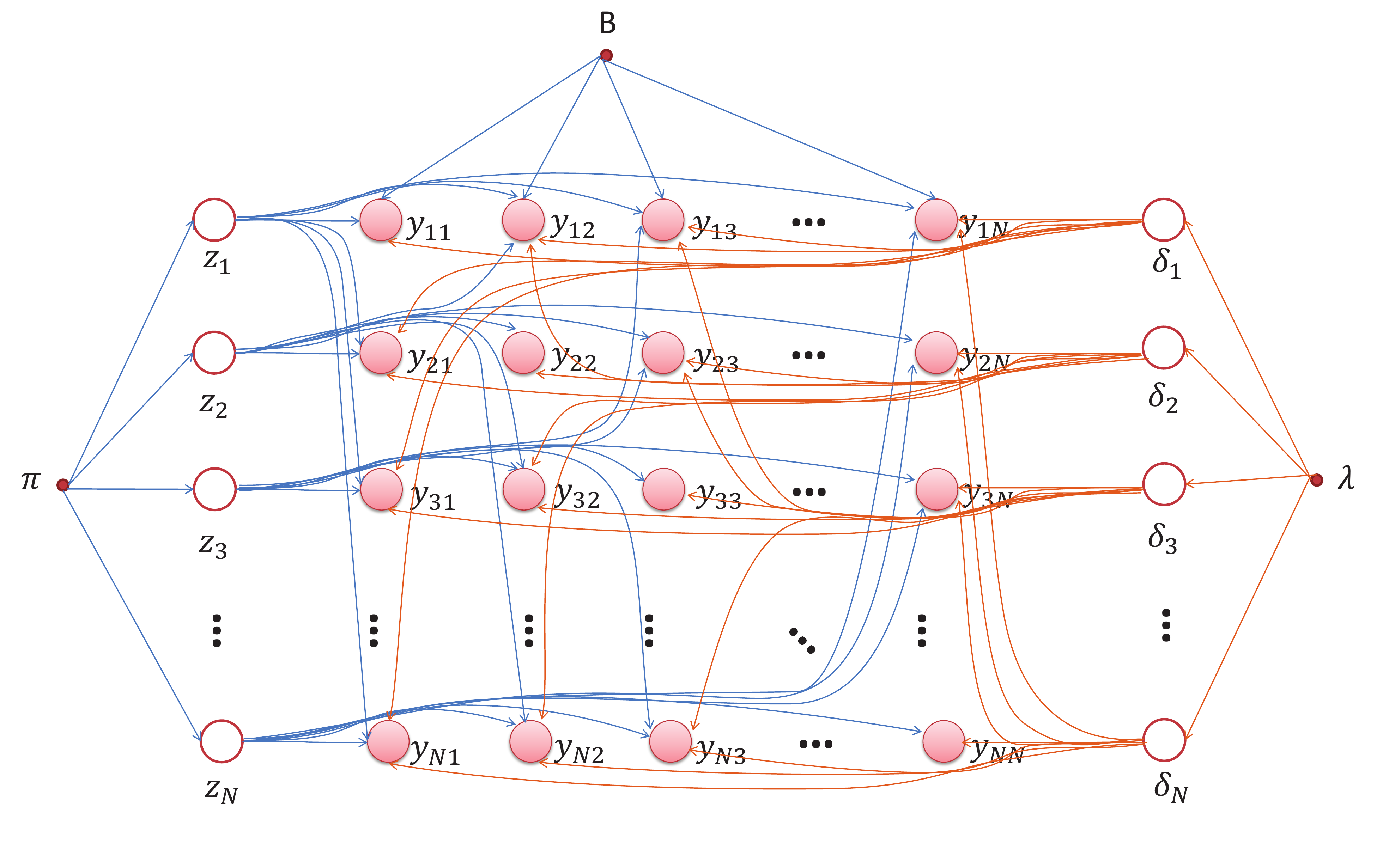}
  \caption{
Graphical representation for PLD-SBM.
 }\label{fig:graphrepresentation}
\end{figure*}

The graph representation for SBM is shown in the left part of Figure \ref{fig:graphrepresentation}, comprised of parameters $(\pi,B)$ and variables $(Y,z_1,\dots,z_N)$. The shadow circles represent the observed edges, while the hollow circles represent the latent cluster index variables. $\pi$ parameterizes the node portion of clusters in the network while $B$ is the cluster-level connectivity parameters. (All observed edges are generatively controlled by $B$. Here, for clarity, we link $B$ just to the first row of adjacency matrix observations.) Clearly, the edges in $Y$ are independent conditioning on cluster indexes $\mathbf z$ \cite{bishop2006pattern}.

Although SBM is generally more able to model directed graphs, undirected graphs, and mixed-membership cluster structures \cite{snijders1997estimation}\cite{airoldi2008mixed}\cite{gopalan2012scalable}\cite{rossi2015role}\cite{wang2014neiwalk}, we have restricted this paper to a very specific setting - undirected graphs with single memberships - to avoid distractions and emphasize our main contribution. Extensions to general cases are, however, rather straightforward.

\subsection{The Generative Model}

Real-world networks often have power-law characteristic over node degrees, where a few nodes of a cluster in the associated graph $\mathcal G$ have considerably large degrees while the rest nodes of it have relatively small degrees. One example is for liberal and conservative communities in a political blog network \cite{adamic2005political}. In this network, each community of a political party contains both high-degree popular blog nodes and low-degree inactive blog nodes. The another example in the Zachary karate club relationship network shows similar degree configuration. Each club community contains high-degree instructor nodes and low-degree student nodes \cite{leger2014detection}.

However, as discussed above, SBM is unable to model this skew degree feature ubiquitously existing in real-world networks. This is because SBM (\ref{eq:SBMs}) treats, in terms of connectivity, all nodes within a cluster $k$ equally with the same group-level parameters, i.e.,  $b_{kk}, b_{k\cdot}, b_{\cdot, k}$, which obviously ignores node-specific information as proven to be useful in other network clustering processes.

In order to capture a heterogeneous degree distribution, we associate each node $i$ with another latent variable, $\delta_i\ge0$, and use it to adjust the generating probability of connectivity.
Specifically,
  \begin{align}\label{eq:HD-SBMs}
    p(y_{ij}=1|z_i,z_j,\delta_i,\delta_j,B)=b_{z_iz_j}^{1+\delta_i+\delta_j}.
  \end{align}
We term $\delta_i$ the degree decay variable. Clearly, these variables $\delta_i$ and/or $\delta_j$ are negatively associated with the probability of single individual connection between nodes $i$ and $j$. As such, with $\delta_i$ varying among nodes, heterogeneous node degree distribution is easily formed. We assign an exponential prior $\mbox{Exp}(\lambda)$ over $\delta_i$ to capture a diverse value range, i.e.,
\begin{align}
    p(\delta_i|\lambda)=\lambda e^{-\lambda\delta_i}.  \nonumber
\end{align}
The graphical representation for our PLD-SBM is shown in Figure \ref{fig:graphrepresentation}. Again, the edge variables are conditionally independent given all hidden variables $\{z_i, \delta_i\}_{i=1}^N$.

Accordingly, the edge generation procedure by the proposed PLD-SBM is summarized as follows:
\begin{itemize}
    \item For each node $i\in \mathcal N= \{1,2,...N\}$,
    \begin{description}
      \item[-] sample the cluster index $z_i\sim\mbox{Multi}(\pi)$, and
      \item[-] sample the degree decay variable $\delta_i\sim\mbox{Exp}(\lambda)$.
    \end{description}
    \item For each node-pair $(i,j)\in\mathcal N\times\mathcal N$,
    \begin{description}
      \item[-] sample the edge $y_{ij} \sim \mbox{Bern}(b_{z_i z_j}^{1+\delta_i + \delta_j})$.
    \end{description}
\end{itemize}
The joint probability distribution over observable edge variables $Y$ and latent variables $(\mathbf z, \boldsymbol \delta)$ in PLD-SBM is then formulated as
  \begin{align}
    &~p(Y,\mathbf z,\boldsymbol \delta|\pi,\lambda,B) \notag\\
    =& \prod_{i,j}p(y_{ij}|b_{z_i,z_j},\delta_i,\delta_j)\prod_{i}p(\delta_i|\lambda)p(z_i|\pi)\notag\\
    =&\prod_{y_{ij}=1}b_{z_i,z_j}^{1+\delta_i+\delta_j}\prod_{y_{ij}=0}[1-b_{z_i,z_j}^{1+\delta_i+\delta_j}]\prod_i\lambda e^{-\lambda \delta_i}\pi_{z_i}. \notag
  \end{align}
Its corresponding graph representation is given in Figure \ref{fig:graphrepresentation}.
Fitting PLD-SBM to a network or graph $\mathcal G$ is achieved by optimizing model parameters $(\pi,\lambda,B)$ with MLE; and a posterior inference on $(\mathbf z,\boldsymbol \delta)$ unveils structural information hidden in the network. Algorithms for estimation and inference will be developed later in Section \ref{sec:algorithm}, however first both theoretically and empirically we demonstrate that the node degrees of the proposed model do follow a power-law distribution.

\subsection{Degree Characteristic}

Although a direct formulation of scale-free network modeling is considerably difficult, intuitive generation procedures that result in a scale-free network have been proposed. For example, in the BA model \cite{barabasi1999emergence}, a network starts from a small number of nodes and grows by each time linking a new node to a fixed number of already presented nodes with connection preference. The connection probability is proportional to the degree of old nodes. However, these scale-free network models make statistical inference inherently difficult, especially when the intention is to incorporate cluster structure modeling as well. Conversely, the state-of-the-art statistical models for network data, e.g., SBM, deviate far from the scale-free characteristic of degree distributions. The proposed PLD-SBM tries to reduce such gaps in the statistical modeling of network. Yet, it should be noted that we are not proving PLD-SBM to be a ``rigorous'' scale-free model, rather that it can improve SBM by incorporating it to power-law degree distributions.

Power-law degree distributions with long heavy tails have been observed on network data in various fields, e.g., biology, social science and Internet studies; and it has been shown that, for many of these networks, the degree distribution can be well fitted by a power-law \cite{barabasi1999emergence}. Let $p(d)$ be the degree distribution of a network, its power-law characteristic is defined by
   \begin{align}\label{eq:power-law0}
    p(d=k)\propto k^{-\gamma},
   \end{align}
with $\gamma$ a shape parameter. As (\ref{eq:power-law0}) is invariant to the scale transformation of a network, such characteristic is also regarded as scale-free.

\begin{figure}[t]
    \centering
    \includegraphics[width=1\columnwidth]{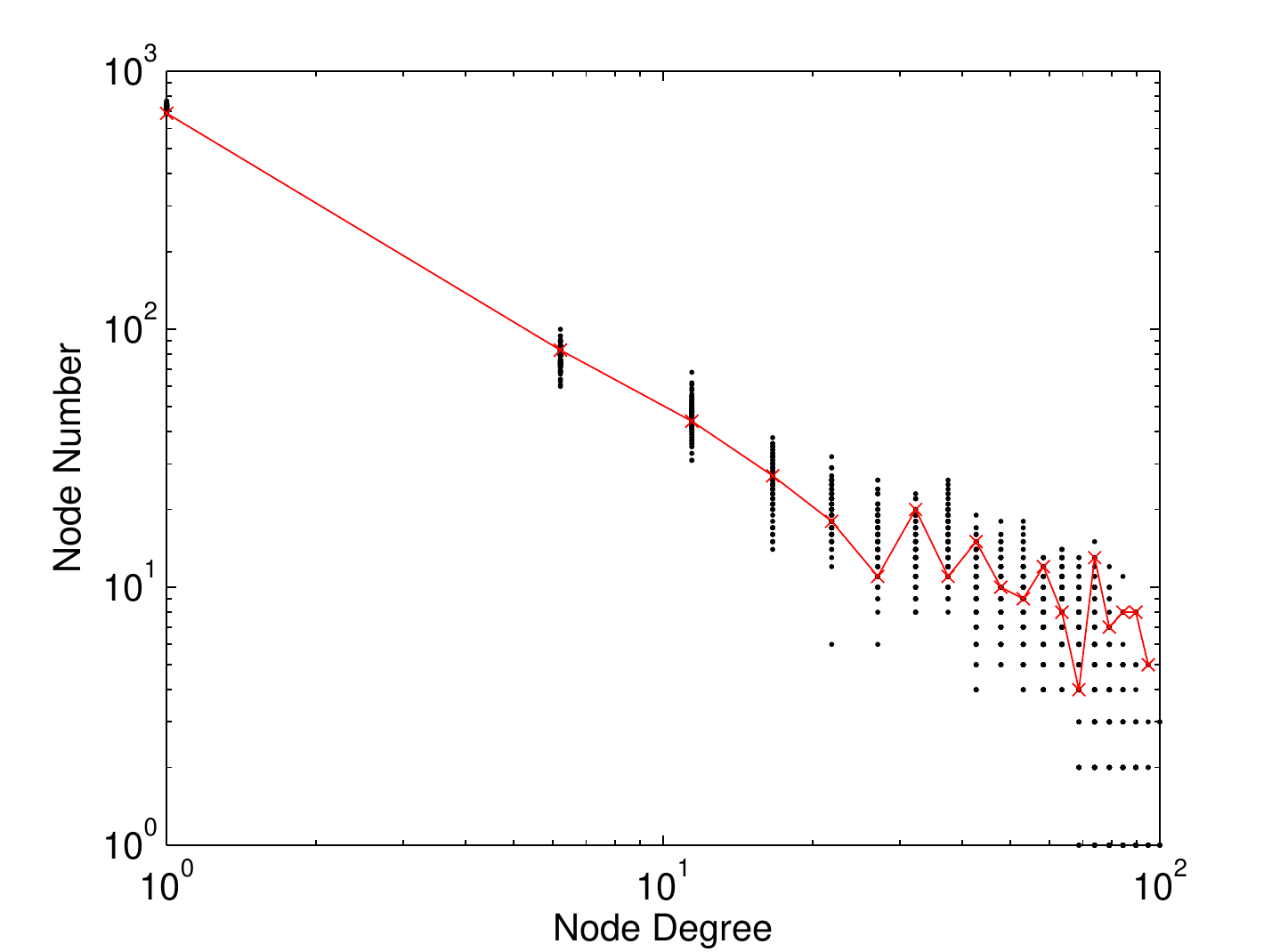}
    \caption{The scale-free characteristic of PLD-SBM. Black dots show the results over 100 simulations, while the red line shows the result from 1 simulation. The best line fitting to the results by least squares has a slope about $-0.892$. }\label{fig:power-law}
\end{figure}

In assortative networks, intra-cluster edges contributes most node degrees. Based on this basic assumption, we consider an intra-cluster or equivalently a single-cluster case to justify the ability of PLD-SBM for degree modeling. With a cluster of $n_0$ nodes, as established by the proposed model, each node is associated with a latent degree decay variable $\delta_i\sim\mbox{Exp}(\lambda)$. Suppose the edge probability between two nodes is $p_0$. Then, based on the Strong Law of Large Numbers (SLLN), as $n_0$ increases, it can be shown that with PLD-SBM the normalized degree of node $i$ will converge to a random variable $\bar d_i$ that only depends on $\delta_i$. Formally,
 \begin{align}\label{eq:expectation}
   \frac1{n_0-1}\sum_{j\ne i}y_{ij}\overset{a.s.}{\longrightarrow} \bar d_i(\delta_i)=\frac{\lambda}{\lambda-\ln p_0}p_0^{1+\delta_i}.
  \end{align}
Applying the fact that $\delta_i$ follows an exponential prior, the distribution of $\bar d_i$ s obeys a power-law
  \begin{align}\label{eq:power_law1}
    p(\bar d_i=k)
     &\propto k^{-(1+\frac\lambda{\ln p_0})},
  \end{align}
where $\gamma = 1+\lambda/{\ln p_0}$ denoting its shape parameter. We term (\ref{eq:power_law1}) as PLD-SBM's power-law degree characteristic. The proof for both (\ref{eq:expectation}) and (\ref{eq:power_law1}) are given in Supplementary. When $\lambda$ is small, the value of the shape parameter $\gamma = 1+\lambda/{\ln p_0}$ in (\ref{eq:power_law1}) approaches $1$. This is smaller than the typical value for real networks, lying between $2$ and $4$. However, smaller shape parameters enable PLD-SBM to adapt to much severer heavy-tail cases in a prior manner.

Note that the power-law degree characteristic (\ref{eq:power_law1}) of PLD-SBM is only for the degree distribution of an individual node, rather than statistic overall as in (\ref{eq:power-law0}). However, in a sparse network (valid for most real networks), $\bar d_i$, $i=1,2,...,N$, are nearly independent, which makes the overall degree distribution (a statistic over i.i.d. examples) similar to that of an individual one, i.e., (\ref{eq:power_law1}). We demonstrate this by a simulation experiment. The setting is: For the exponential distribution parameter, $\lambda=0.01$, and the probability for linking any node pairs, $p_0=0.9$. Then, $100$ single-cluster networks of $n_0=1000$ nodes are generated. Empirical degree distribution, i.e., node number vs node degree, is shown in Figure \ref{fig:power-law}. Clearly, it concentrates approximately along a line in the log-domain, with a slope about $-0.892$. When comparing to (\ref{eq:power_law1}), the slope is $ - (1+\lambda/{\ln p_0}) = -0.905$ - very close to the simulation result. We refer interesting readers for better power-law fitting methods to \cite{clauset2009power}. We conclude that the networks generated by PLD-SBM do follow a power-law overall.

\section{EM Algorithm} \label{sec:algorithm}

In this section, we derive a Viterbi-type variational EM algorithm \cite{yu2016distribution}\cite{xuan2017doubly} for the implementation of PLD-SBM.

\subsection{Viterbi-type E-Step}

Given observations of the adjacency matrix $Y$ and model parameters $(\pi,\lambda,B)$, the posterior distribution of latent variables $(\mathbf z,\boldsymbol \delta)$ is required to reveal the underlying cluster structure. However, this joint posterior $p(\mathbf z,\boldsymbol \delta|Y,\pi,\lambda,B)$ has no closed-form, because the calculation of observation likelihood $p(Y|\pi,\lambda,B)$ requires integrals over all latent variables which are generally analytically intractable. Instead, variational methods \cite{lu2016bayesian} provide a tractable way via approximating the true posterior with a certain distribution family that usually can be learned efficiently. The mean-field method is within this category, and applies a fully factorized distribution family based on the assumption that all associated variables are independent to each other. For SBM, it is
  \begin{align}\label{eq:variational_posterior}
   q(\mathbf z,\boldsymbol \delta)=\prod_{i}q(z_i)q(\delta_i).
  \end{align}
Generally, the form of variational posterior distributions $q(z_i)$ and $q(\delta_i)$ is chosen, based on computational convenience, from the same family of prior distributions to take advantage of the conjugate between likelihood and prior. However, this is not the case in our problem. Instead, we set
  \begin{align}\label{eq:var-distribution}
    q(z_i)=\mbox{Multi}(\phi_i)~\mbox{and}~ q(\delta_i) = \mathbf 1 (\bar{\delta_i}),
  \end{align}
where $q(z_i)$ is a multinomial distribution parameterized with $\phi_i$, while $q(\delta_i)$ is a degenerated distribution with probability 1 at point $\bar{\delta_i}$.

Here, the use of degenerated distribution $q(\delta_i)$ is inspired by the Viterbi-type EM algorithm used for training Hidden Markov Models (HMM) \cite{juang1990segmental}, and is based on whether it should be concentrated somewhere in the positive half real line in the posterior sense. .

One optimal set of variational distributions is to maximize the objective of marginal likelihood of observation and is obtained via approximating the true posterior within family (\ref{eq:variational_posterior}). Specifically, it is attained by minimizing the Kullback-Leibler (KL) divergence between $q(\mathbf z,\boldsymbol \delta)$ and $p(\mathbf z, \boldsymbol \delta|Y,\pi,\lambda,B)$. Often, minimization is equivalently transformed to the maximization of an evidence lower bound (ELBO) of the marginal likelihood of observed edges. By applying Jensen's inequality \cite{Jordan:1999:IVM:339248.339252} to the marginal likelihood, the ELBO $\mathcal L(\boldsymbol \phi, \boldsymbol \delta)$ for our problem is given by
  \begin{align}%\label{eq:elbo}
    &\log p(Y|\pi,\lambda,B)\ge\notag\\
    &\quad\mathcal L(\boldsymbol \phi,\boldsymbol \delta)
    =\mathbb E_q\log p(Y,\mathbf z,\boldsymbol \delta|\pi,\lambda,B) - \mathbb E_q \log q(\mathbf z,\boldsymbol \delta) \notag
  \end{align}
  By (\ref{eq:var-distribution}) and Taylor approximation \cite{wang2013variational}\cite{ahmed2007tight}, we get:
  \begin{align}\label{eq:elbo_appro}
    \mathcal L(\boldsymbol \phi,\boldsymbol \delta)
    &\approx\sum_{y_{ij}=1}(1+\bar \delta_i+ \bar \delta_j)\sum_{k}\sum_{k'}\phi_{ik}\phi_{jk'}\log b_{kk'} \notag\\
    &-\sum_{y_{ij}=0}\sum_{k}\sum_{k'}\phi_{ik}\phi_{jk'}b_{kk'}^{1+\bar \delta_i+\bar \delta_j}\notag\\
    &+\sum_{i}\sum_{k}\phi_{ik}\log\pi_{k}-\sum_{i}\sum_{k}\phi_{ik}\log\phi_{ik}\notag\\
    &+N\log\lambda -\lambda \sum_i \bar \delta_i.
  \end{align}
Here for networks of sparse and with assortative communities, which is the situation we are concerned, $b_{kk'}$s are not large, and a small portion of $delta_i$s are rather large. Under these cases, this Taylor approximation is valid.

Although (\ref{eq:elbo_appro}) is not jointly concave w.r.t. $(\boldsymbol \phi,\boldsymbol{\bar \delta})$, it is verifiable that it is concave w.r.t. each individual variable $\phi_i$ and $\bar \delta_i$. Thus we apply coordinate gradient ascend to alternatively optimize these variables.
We present the results here, and the detailed derivations are described in Supplementary.
The associated gradient for $\bar \delta_i$ is computed by
 \begin{align} \label{eq:delta-grad}
  \frac{\partial \mathcal{O}(\bar \delta_i)}{\partial \bar \delta_i} &=
       \sum_{k}\phi_{ik}\sum_{k'}\log b_{kk'}\sum_{y_{ij}=1}\phi_{jk'} \notag \\
       &- \sum_{k}\phi_{ik}\sum_{y_{ij}=0}\sum_{k'}\phi_{jk'}b_{kk'}^{1+\bar \delta_i+\bar \delta_j} \ln (b_{kk'})
       -\lambda,
  \end{align}
and the updating for $\phi_i$ is as,
  \begin{align}\label{eq:opt-phi}
    \phi_{ik}\propto &\pi_k\exp\left\{\sum_{y_{ij}=1}\sum_{k'}(1+\bar \delta_i+ \bar \delta_j)\phi_{jk'}\log b_{kk'}\right.\notag\\
    &\qquad\qquad\left.-\sum_{y_{ij}=0}\sum_{k'}\phi_{jk'}b_{kk'}^{1+\bar \delta_i + \bar \delta_j}\right\}.
  \end{align}
From (\ref{eq:opt-phi}), when optimizing $\phi_i$, i.e., inferring the cluster membership of node $i$, the membership and degree decay of all other nodes matter. The two terms in the $\mbox{exp}(\cdot)$ operator count respectively to contribution of linked nodes and unlinked nodes, but in different manner. When $y_{ij}=1$, i.e., node $j$ is linked to node $i$, its membership is considered with extra weight extra weight $\bar \delta_j$ due to $\bar \delta_j\ge 0$ and $b_{kk'} \le 1$. By contrast, when $y_{ij}=0$, less weight is taken because $b_{kk'}^{1+\bar \delta_i+ \bar \delta_j}$ decreases for large $\bar \delta_j$. This differs PLD-SBM from SBM, which treats the contribution of each node $j$ uniformly, and thus helps correct inference bias induced by power-law degree distributions.

\subsection{M-Step}

When it comes to the optimization of model parameters $(\pi,\lambda,B)$, ELBO (\ref{eq:elbo_appro}) is again applied as the objective, but with the entropy term of $q$ dropped since it is irrelevant to model parameters. First, we set a fixed configuration $\lambda=0.01$ for practical use (We tried different values from $\{0.01,0.1,1,10\}$ for $\lambda$, but no performance difference is observed. Therefore, we simply fixed $\lambda$ as $0.01$ following that small $\lambda$ leading to small shape parameters which enables PLD-SBM to adapt to much severer heavy-tail cases). Then, the optimal $B$ is obtained iteratively by gradient ascend method, and the associated gradient is
 \begin{align} \label{eq:b-grad}
  \frac{\partial \mathcal O(b_{kk'})}{\partial b_{kk'}} &= \frac{\sum_{y_{ij}=1}(1+\bar \delta_i+\bar \delta_j)\phi_{ik}\phi_{jk'} }{b_{kk'}}  \notag \\
  & - \sum_{y_{ij}=0} (1+\bar \delta_i + \bar \delta_j) \phi_{ik}\phi_{jk'}b_{kk'}^{\bar \delta_i+\bar\delta_j}
  \end{align}
Finally, the optimal $\pi$ has closed-form, and each unnormalized element is computed by
  \begin{align} \label{eq:update-pi}
   \pi_k\propto\sum_{i}\phi_{ik}.
  \end{align}
The overall pseudo code of above algorithm is summarized in Algorithm \ref{alg:viterbi-typeEM}.

\begin{algorithm}[!tp]
\caption{Inference for PLD-SBM}
\begin{algorithmic}[1]
\REQUIRE{Initialization for model parameters $\lambda, B, \pi$ and variational parameters $\boldsymbol \phi,\boldsymbol{\bar \delta}$;
the number of communities $K$;
stop criterion $\varepsilon$}.
\STATE Compute variational likelihood $\mathcal L^{new}$ by (\ref{eq:elbo_appro}).
\REPEAT
\STATE $\mathcal L^{old} = \mathcal L^{new}$.
\STATE \textbf{variational E-step}
\STATE update $\boldsymbol{\bar \delta}$ by coordinate gradient ascend and gradient is given by (\ref{eq:delta-grad}).
\STATE update $\boldsymbol \phi$ via (\ref{eq:opt-phi}).
\STATE \textbf{M-step}
\STATE update $B$ by coordinate gradient ascend and gradient is given by (\ref{eq:b-grad}).
\STATE update $\pi$ via (\ref{eq:update-pi}).
\STATE Compute variational likelihood $\mathcal L^{new}$ with updated parameters by (\ref{eq:elbo_appro})
\UNTIL {$|\mathcal L_A^{new} - \mathcal L_A^{old}|<\varepsilon$}
\STATE predict the cluster index $z_i$ of each node via (\ref{eq:max-cluster}).
\RETURN learned model parameters $\boldsymbol{\bar \delta}^*, \pi^*, B^*$ and cluster structure $\boldsymbol z^*$
\end{algorithmic}
\label{alg:viterbi-typeEM}
\end{algorithm}

\begin{figure}[tb]
  \centering
  \includegraphics[width=1\columnwidth]{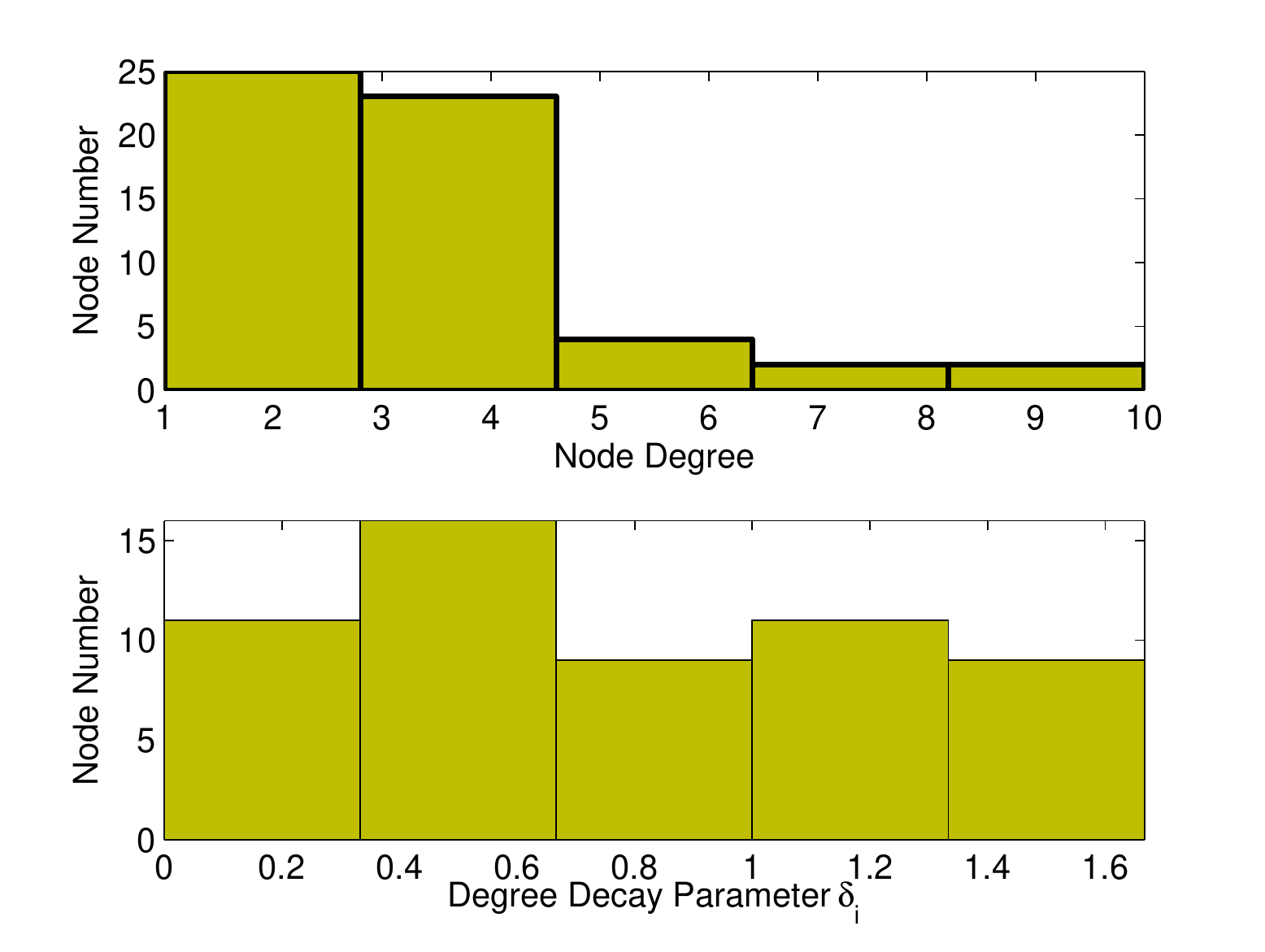}
  \caption{Histograms of Node Degrees and Estimated Degree Decay Parameters $\bar \delta_i$ on one Simulated Network.}\label{fig:degree_decay_simulation}
\end{figure}

\subsection{Algorithm Complexity Analysis}

The computation complexity is roughly analyzed.
Line 4 in Algorithm \ref{alg:viterbi-typeEM} updates $\{\delta_i\}_{i=1,\dots, N}$, and its complexity is
${O}(N\times \max\{N_{i,\neg E}\}_{i=1,\dots,N} \times K^2)$ with $N$ the number of network nodes, $K$ the number of clusters, and $N_{i,\neg E}$ equal to $N$ minus the degree of node $i$.
Line 5 updates $\boldsymbol \phi$, where each element takes ${O}(\max\{N_{i,\neg E},N_{i,E}\}\times K)$ and its total time complexity is ${O}(N\times \max\{N_{i,\neg E},N_{i,E}\}\times K^2)$, with $N_{i,E}$ the degree of node $i$.
Line 7
updates $B$ with time complexity of $O(K^2\times \max\{N_{i,\neg E}, N_{i,E}\}_{i=1,\dots,N})$, and Line 9 calculates the likelihood which takes the highest time complexity of $O(\max\{N_{\neg E}, N_{E}\} \times K^2)$ with $N_{\neg E}$ the number of non-edge pairs and $N_E$ the number of edges. Suppose the number of EM iterations for convergence is $T$, then the overall computation complexity is $O(T \times \max\{N_{\neg E}, N_{E}\} \times K^2)$.

Accordingly, the complexity of our algorithm is of order $O(N^2)$, as both $y_{ij}=1$ and $y_{ij}=0$ are used in calculating the objective. The same complexity is shared by SBM in the literature. As most real networks are sparse, i.e., the edge number $N_{E}$ is much less than $N(N-1)/2$. Thus, it is common to speedup the algorithm by sampling a subset of non-edges, i.e., the entries $y_{ij}=0$, to get an approximate estimate of the objective \cite{gopalan2012scalable}. When the size of the non-edge subset is less than $N_E$, the algorithm complexity is reduce to $O(T\times N_E \times K^2)$.

\subsection{Choice of the number of communities}

The number of network communities is unknown in practice. How to estimate it under SBM has been extensively studied \cite{yan2014model}\cite{latouche2014model}\cite{wang2015likelihood}\cite{newman2016estimating}. However, as this issue is not our main concern, we simply adopt the criterion of integrated complete log-likelihood (ICL) developed by \cite{biernacki2000assessing}\cite{daudin2008mixture} to choose the community numbers for different experimental datasets. In our case the criterion is calculated as
\begin{align}
\mathrm{ICL}(K) &= \log p(Y, \boldsymbol z^*, \boldsymbol{\bar \delta}^*|K, \pi^*, B^*) \notag \\
&- \frac{K-1}{2}\log N
- \frac{1}{2} \frac{K(K+1)}{2}\log \frac{N(N-1)}{2}, \notag
\end{align}
where $\boldsymbol z^*, \boldsymbol{\bar \delta}^*, \pi^*, B^*$ are the outputs of the proposed learning and inference procedure. Here, the first term denotes the fitness of the model with $K$ communities, while the second and third terms penalize the model complexity.

\section{Simulation Studies}\label{sec:simulations}

In this section, we evaluate the performance of PLD-SBM on simulated networks either biased to SBM or biased to PLD-SBM. Since the cluster structure is known during the data generation process, we are able to  exactly measure the performance of node clustering, and show how PLD-SBM improves SBM.

\subsection{Networks with Homogeneous Degree Distribution}

\begin{figure}[tb]
  \centering
  \begin{subfigure}[b]{0.24\textwidth}
                \includegraphics[width=1\textwidth]{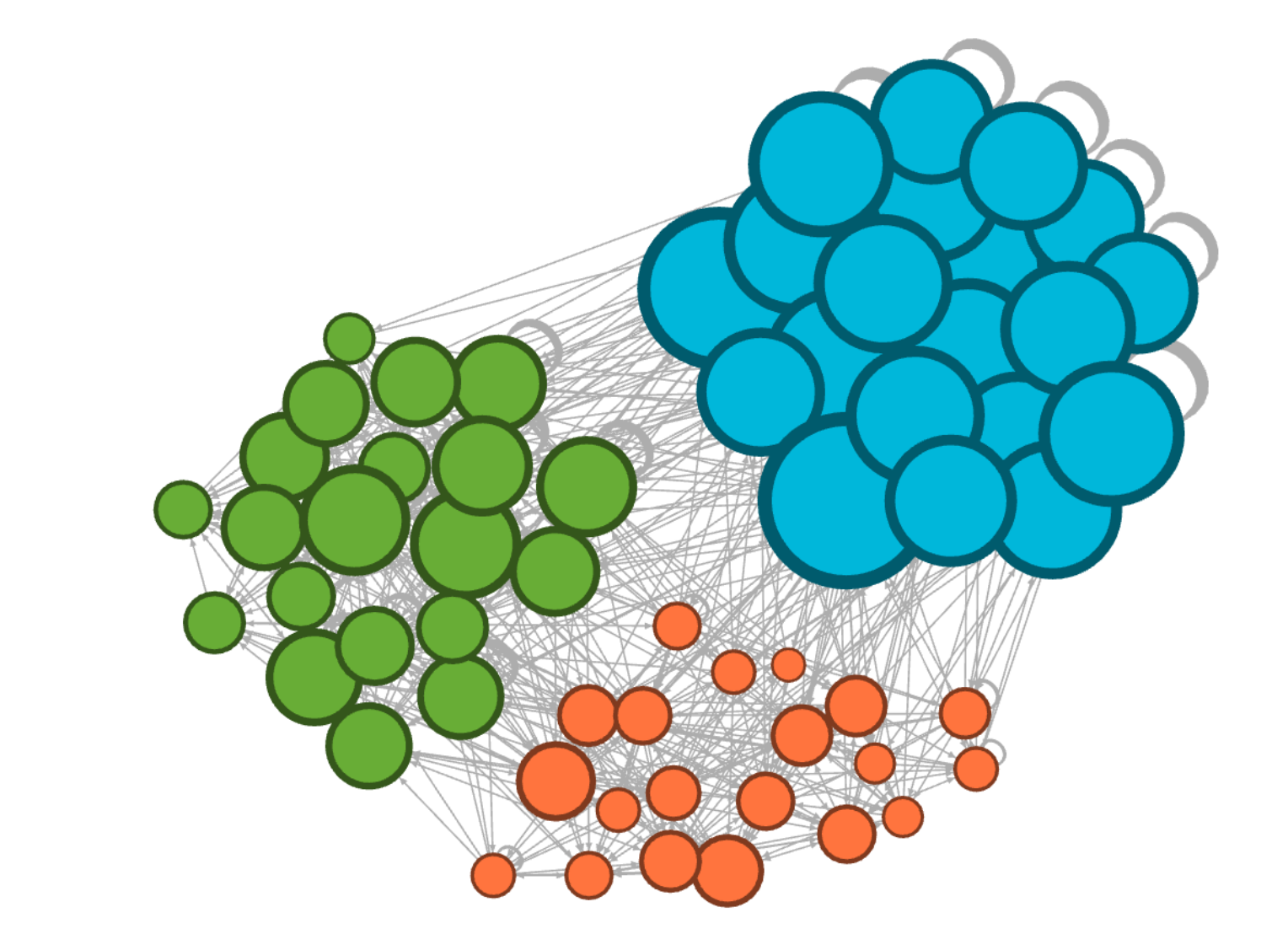}
                \caption{Truth}
                \label{fig:SBMTruth}
  \end{subfigure}
  \begin{subfigure}[b]{0.24\textwidth}
                \includegraphics[width=1\textwidth]{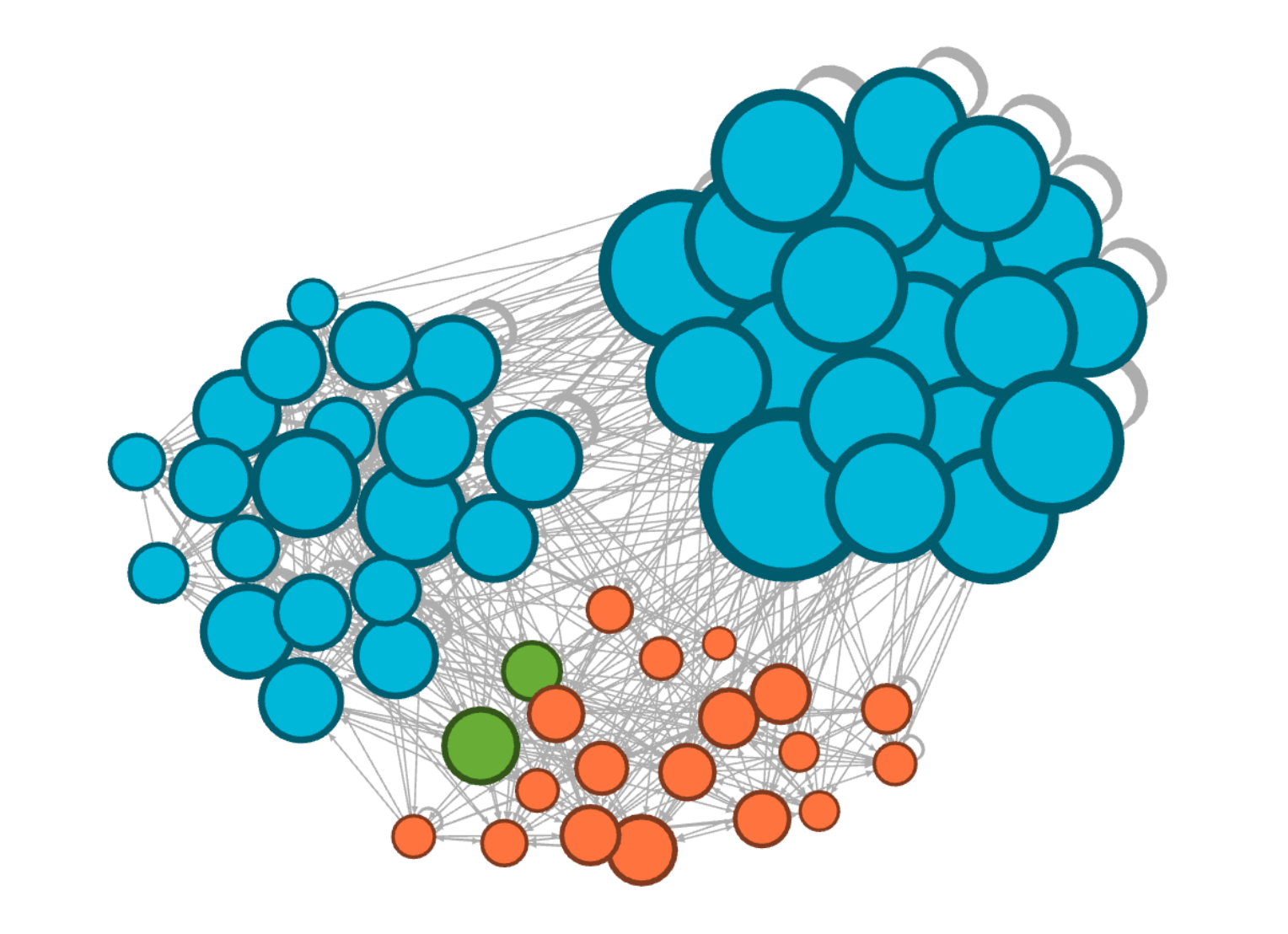}
                \caption{SBM}
                \label{fig:SBMSBM}
  \end{subfigure}%
  \\
   \begin{subfigure}[b]{0.24\textwidth}
                \includegraphics[width=1\textwidth]{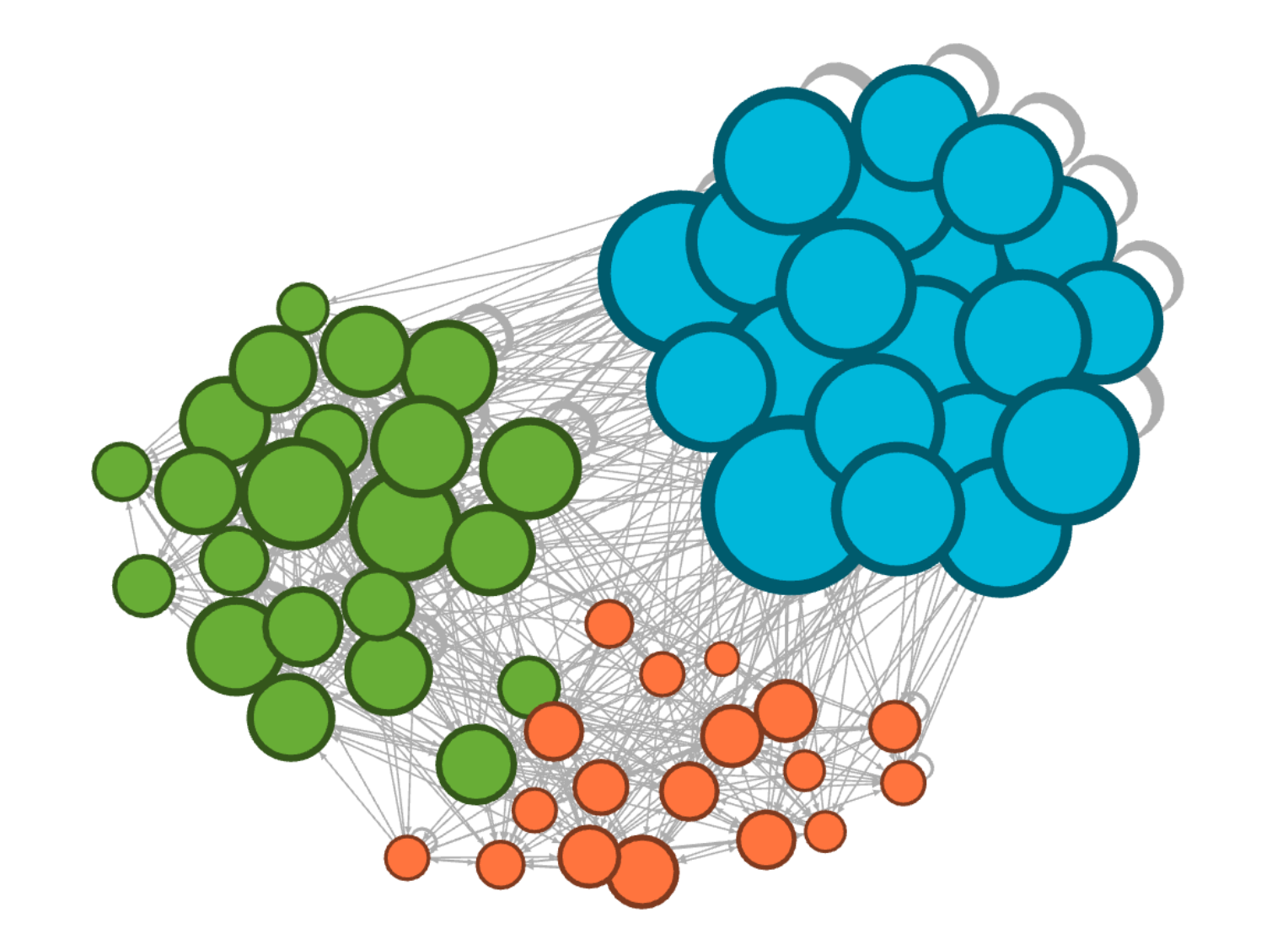}
                \caption{PLD-SBM}
                \label{fig:SBMPLDSBM}
  \end{subfigure}%
   \begin{subfigure}[b]{0.24\textwidth}
                \includegraphics[width=1\textwidth]{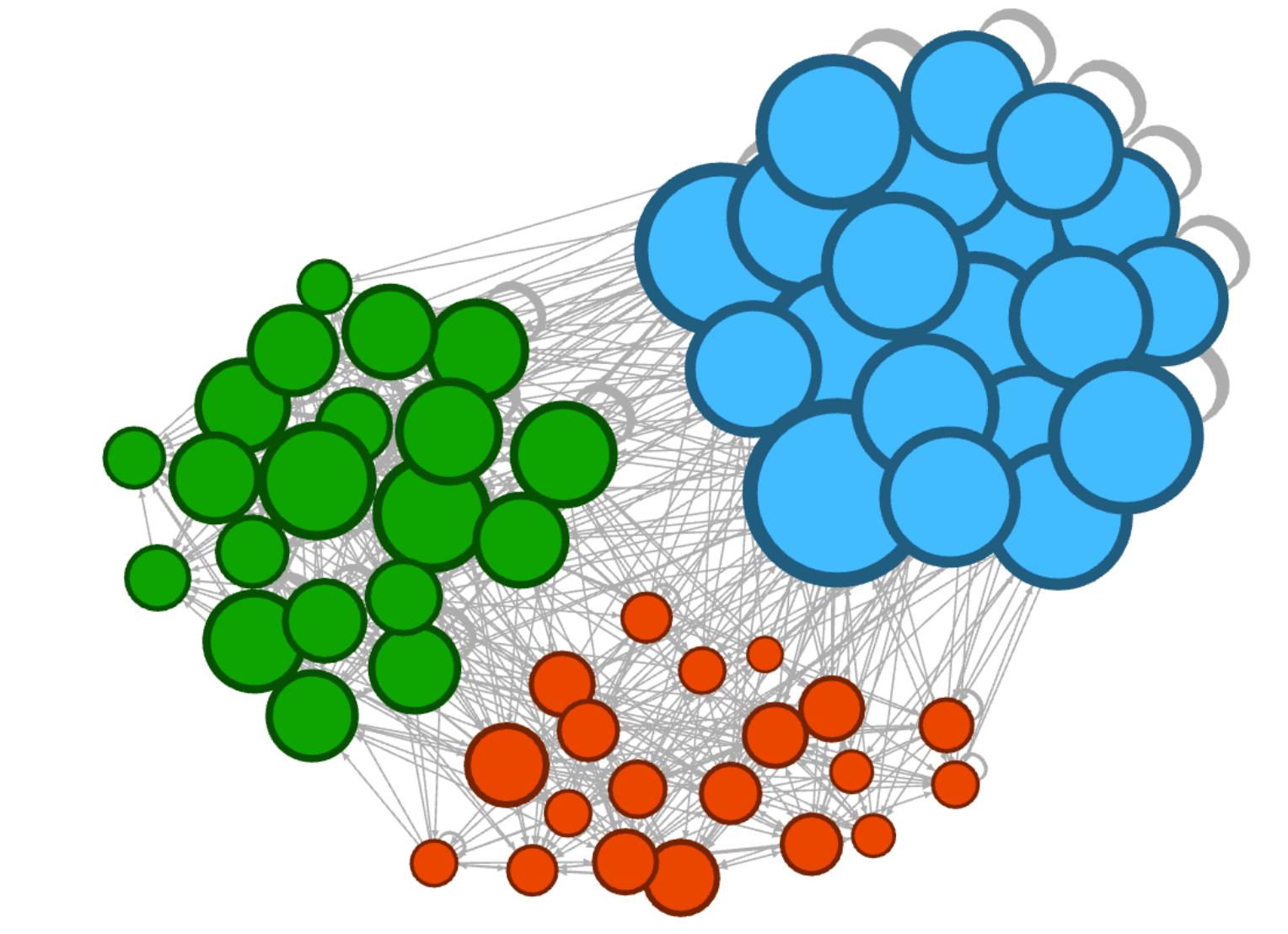}
                \caption{DC-SBM}
                \label{fig:SBMDCSBM}
  \end{subfigure}%
  \caption{Clustering results on SBM-generated network.}\label{fig:SBMGenerating}
\end{figure}

The network generating setting is: Three clusters of $20$ nodes are generated with different intra-cluster link probabilities, i.e., $0.3$, $0.7$, and $0.9$; the $\lambda$ is fixed to $0.1$; all inter-cluster link probabilities are the same and equal to $0.1$; all links are generated independently once the cluster labels of the relevant nodes are given \cite{holland1983stochastic}.  One generated network is shown in Figure \ref{fig:SBMTruth}. The size of each node is proportional to its degree. It clearly shows that the degrees are almost uniformly distributed within each community, and do not follow any power-law alike distributions.

SBM, supposed to fit the network best, surprisingly achieves the worst community detection in terms of similarity to the ground truth, as shown in Figure \ref{fig:SBMSBM}. This might because the nodes within the wrongly clustered-together block have quite similar degrees, as visualized with similar circle sizes in the figure. By contrast, both DC-SBM and the proposed PLD-SBM achieves similar and better cluster accuracy. This should attribute to their explicit degree correct consideration.

\subsection{Networks with Heterogeneous Degree Distribution}
The network generating setting is as below. $20$ networks are generated, each with $3$ clusters and each cluster with around $20$ nodes\footnote{Due to the scale-free network model, no fixed but only approximate node numbers are obtained in the resulting network}. The edges between nodes are generated in two steps. First, the BA model \cite{barabasi1999emergence} is applied to generate intra-cluster links. Next, $5$ node pairs from each cluster pair are randomly picked to form inter-cluster edges.

 \begin{figure}[tb]
    \centering
    \includegraphics[width=0.9\columnwidth]{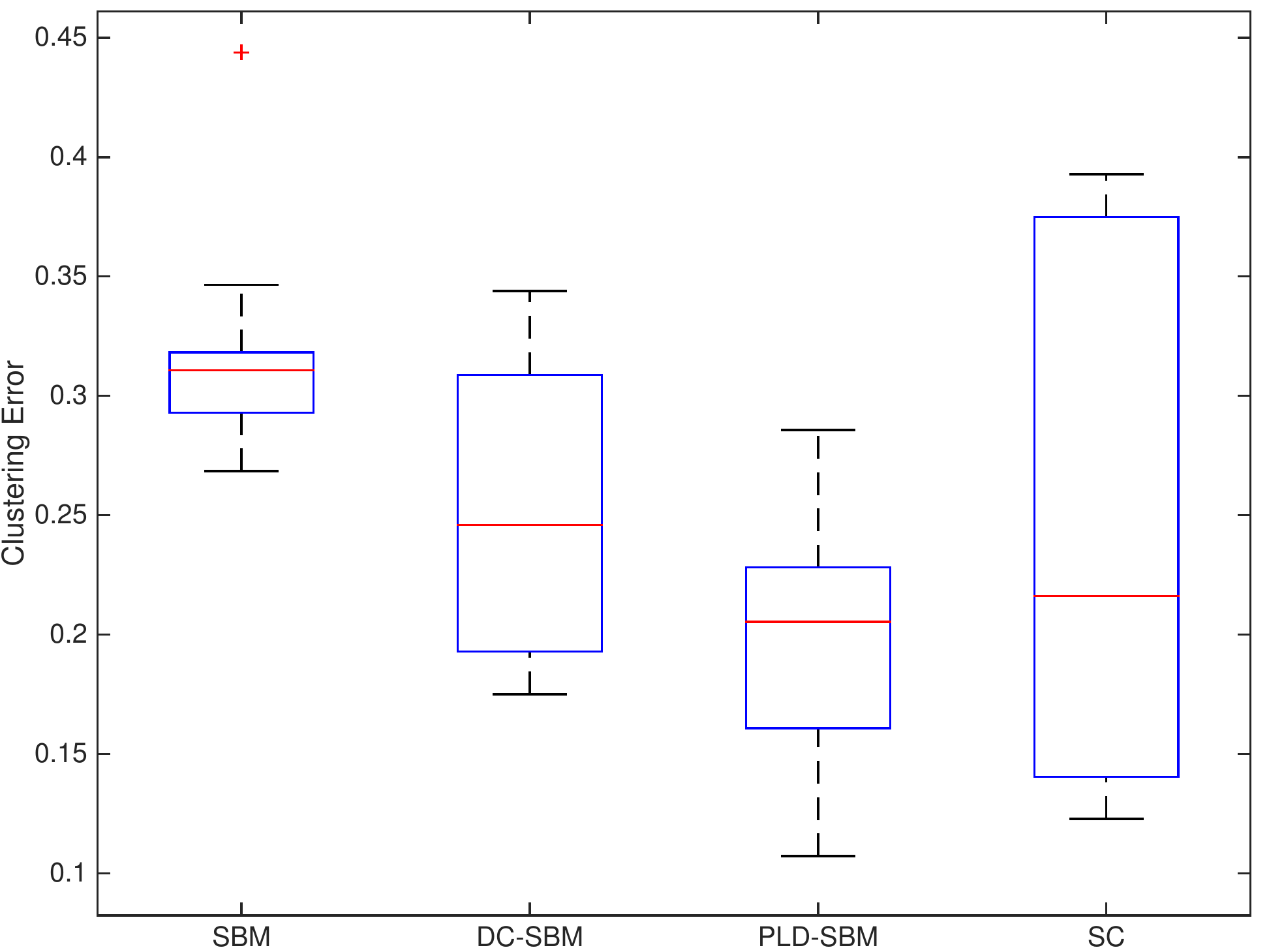}
    \caption{Clustering Error Comparison on Simulated Networks: SBM, stochastic block model\protect\cite{snijders1997estimation}; DC-SBM, degree-corrected SBM; PLD-SBM, power-law degree SBM; SC, spectral clustering \protect\cite{Luxburg:2007TutorialSpeClus}.}\label{fig:cluster-perf}
  \end{figure}

   \begin{figure}[tb]
    \centering
    \includegraphics[width=1\columnwidth]{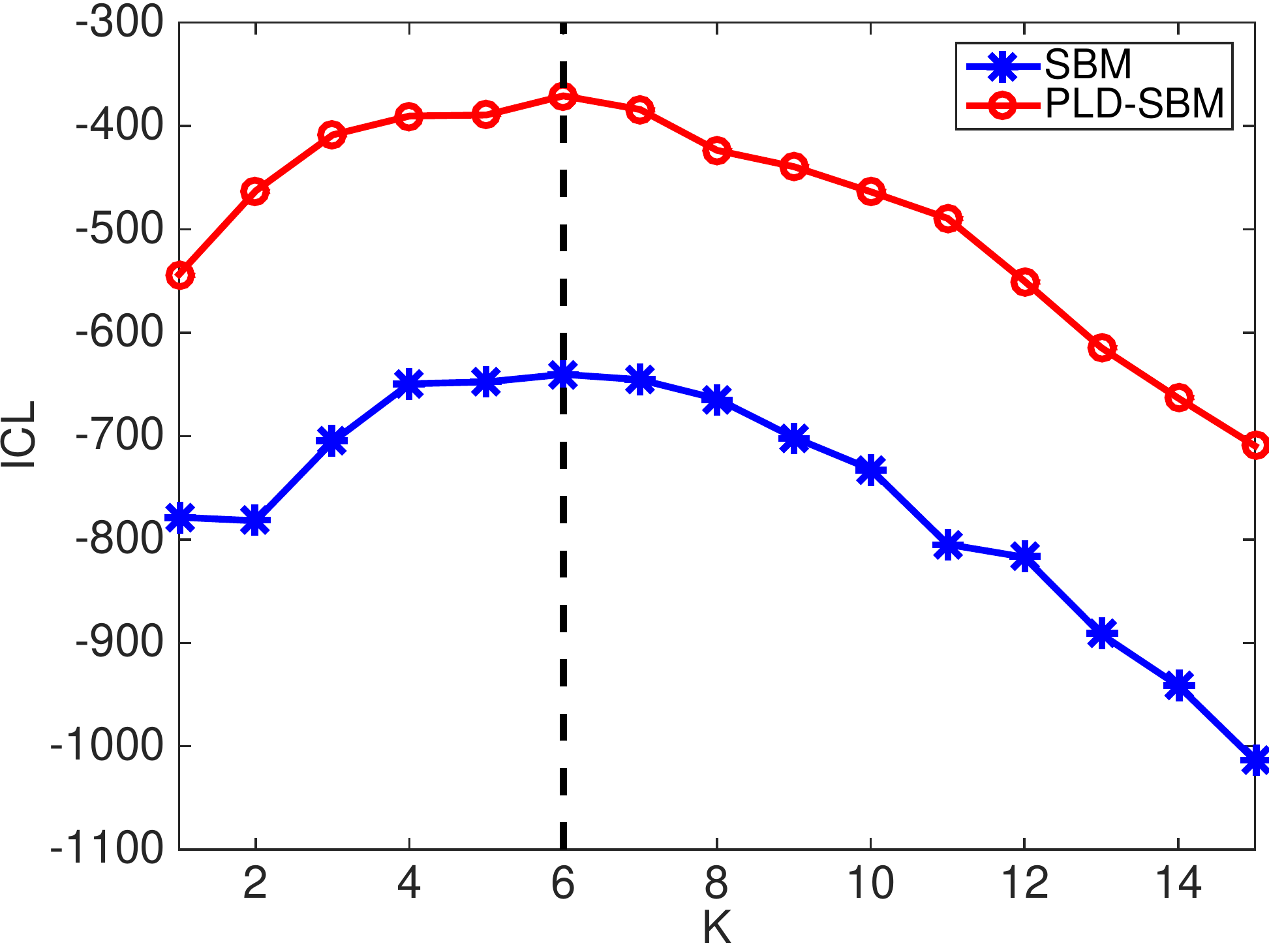}
    \caption{Model Selection of $K$ on Adolescent Health Data. The dashed black vertical line indicates the truth number of communities, i.e., $K=6$. }\label{fig:ahd_K}
  \end{figure}

\begin{table*}
  \centering
  \renewcommand{\arraystretch}{1.8}
  \setlength{\tabcolsep}{6.5pt}
  \begin{tabular}{r|*{6}{c}|*{6}{c}|*{6}{c}|*{6}{c}}
  &\multicolumn{6}{|c|}{SBM Clusters} &\multicolumn{6}{|c|}{PLD-SBM Clusters} &\multicolumn{6}{|c|}{MMSBM Clusters} &\multicolumn{6}{|c}{MSBM Clusters}  \\
  Grade&1&2&3&4&5&6&1&2&3&4&5&6&1&2&3&4&5&6&1&2&3&4&5&6\\
    \hline
    7   &13&1&0&0&0&0    &13&1&0&0&0&0    &13&1&0&0&0&0    &13&1&0&0&0&0\\

    8   &0&11&1&0&0&0    &0&11&1&0&0&0    &0&9&2&0&0&1     &0&10&2&0&0&0\\

    9   &0&0&14&0&0&2    &0&0&14&0&0&2   &0&0&16&0&0&0    &0&0&10&0&0&6\\

    10  &0&0&0&8&0&2     &0&0&0&10&0&0    &0&0&0&10&0&0    &0&0&0&10&0&0\\

    11  &0&0&0&2&7&4     &0&0&0&0&11&2    &0&0&1&0&11&1    &0&0&1&0&11&1\\

    12  &0&0&0&0&1&3     &0&0&0&0&1&3    &0&0&0&0&0&4     &0&0&0&0&0&4\\
    \hline
    &\multicolumn{6}{|r|}{Total Error 13  }&\multicolumn{6}{|r}{Total Error 7  }&\multicolumn{6}{|r|}{Total Error 6  }&\multicolumn{6}{|r|}{Total Error 10  }
  \end{tabular}
  \caption{Performance Evaluation on the Adolescent Health Network: SBM, stochastic block model \protect\cite{snijders1997estimation}; PLD-SBM, power-law degree stochastic block model; MMSBM, mixed membership stochastic block model \protect\cite{airoldi2008mixed}; MSBM, stochastic block mixture model \protect\cite{Doreian07Discussion}.}\label{tab:cluster-perf}
\end{table*}

\begin{figure*}[tb]
  \centering
  \begin{subfigure}[b]{0.5\textwidth}
                \centering
                \includegraphics[width=1\textwidth]{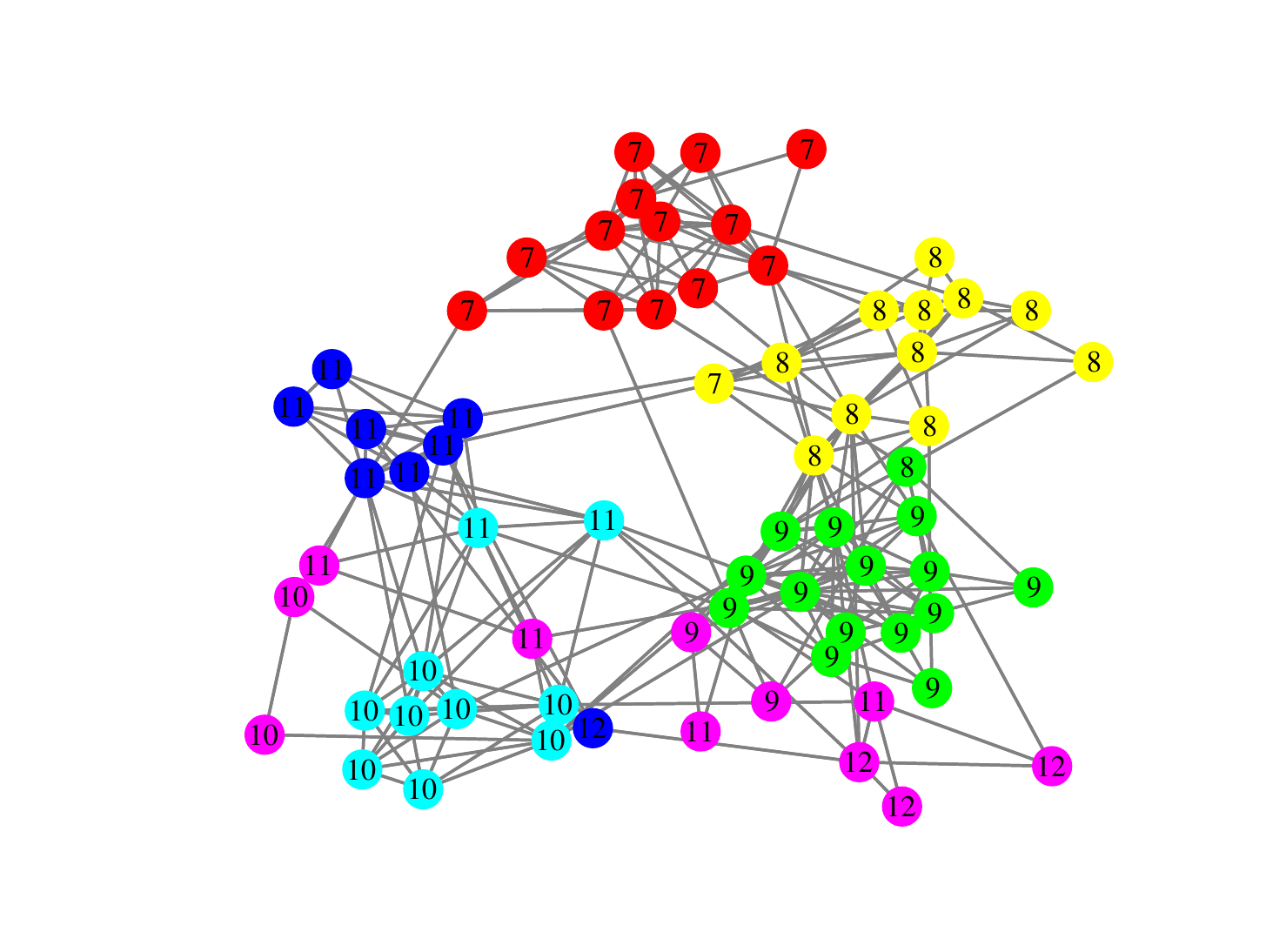} % figure5a
                \caption{SBM}
                \label{fig:SBM}
  \end{subfigure}
\hspace{-3em}
  \begin{subfigure}[b]{0.5\textwidth}
                \centering
                \includegraphics[width=1\textwidth]{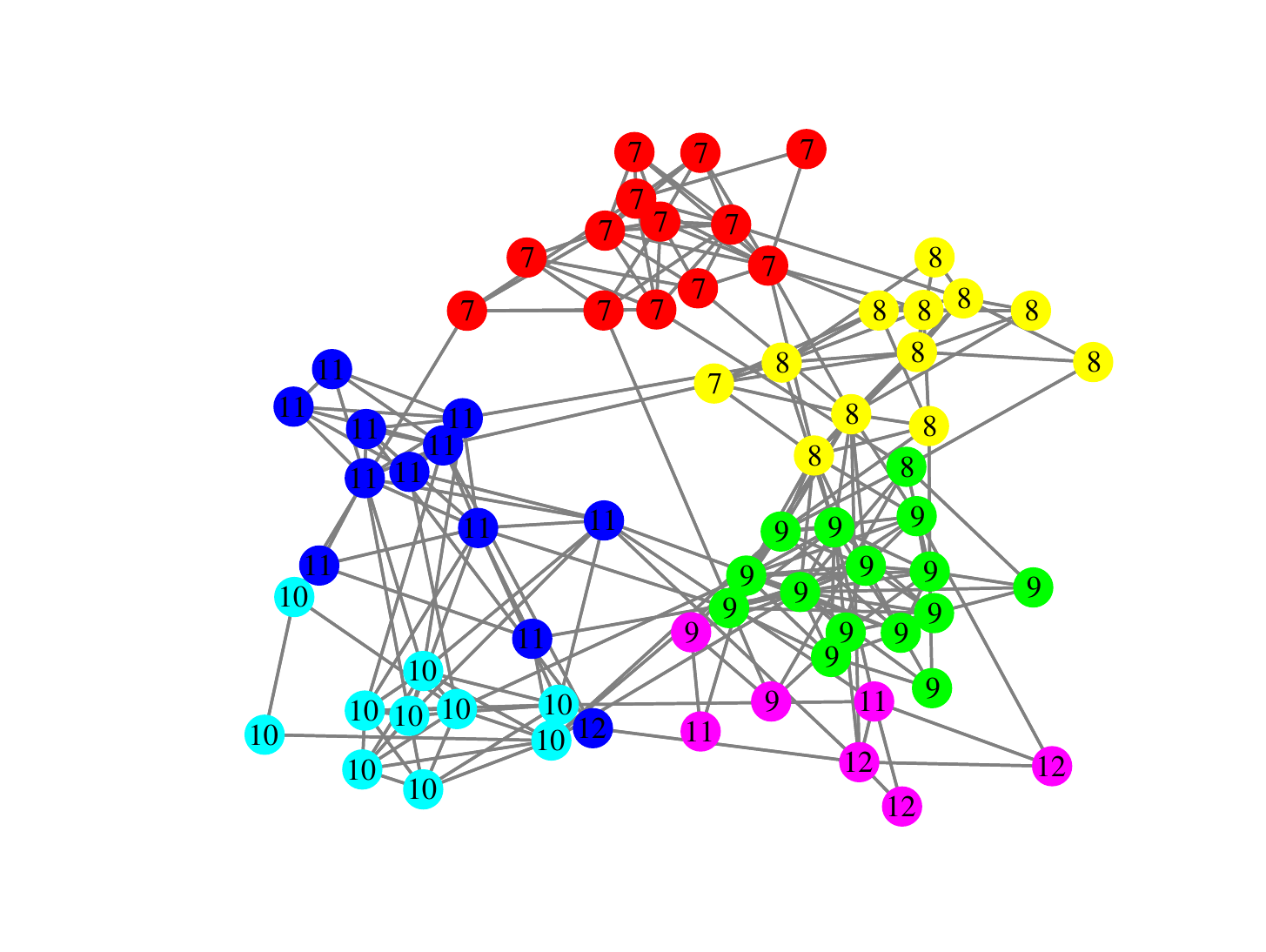} %figure5b
                \caption{PLD-SBM}
                \label{fig:HD-SBM}
  \end{subfigure}%
  \caption{Prediction on the Adolescent Health Data with $K=6$: SBM, stochastic block model \protect\cite{snijders1997estimation}; PLD-SBM, power-law degree stochastic block model. Color stands for predicted cluster; number indicates true grade.}\label{fig:visual_add_health}
\end{figure*}

\begin{figure}[tb]
  \centering
  \includegraphics[width=1.0\columnwidth]{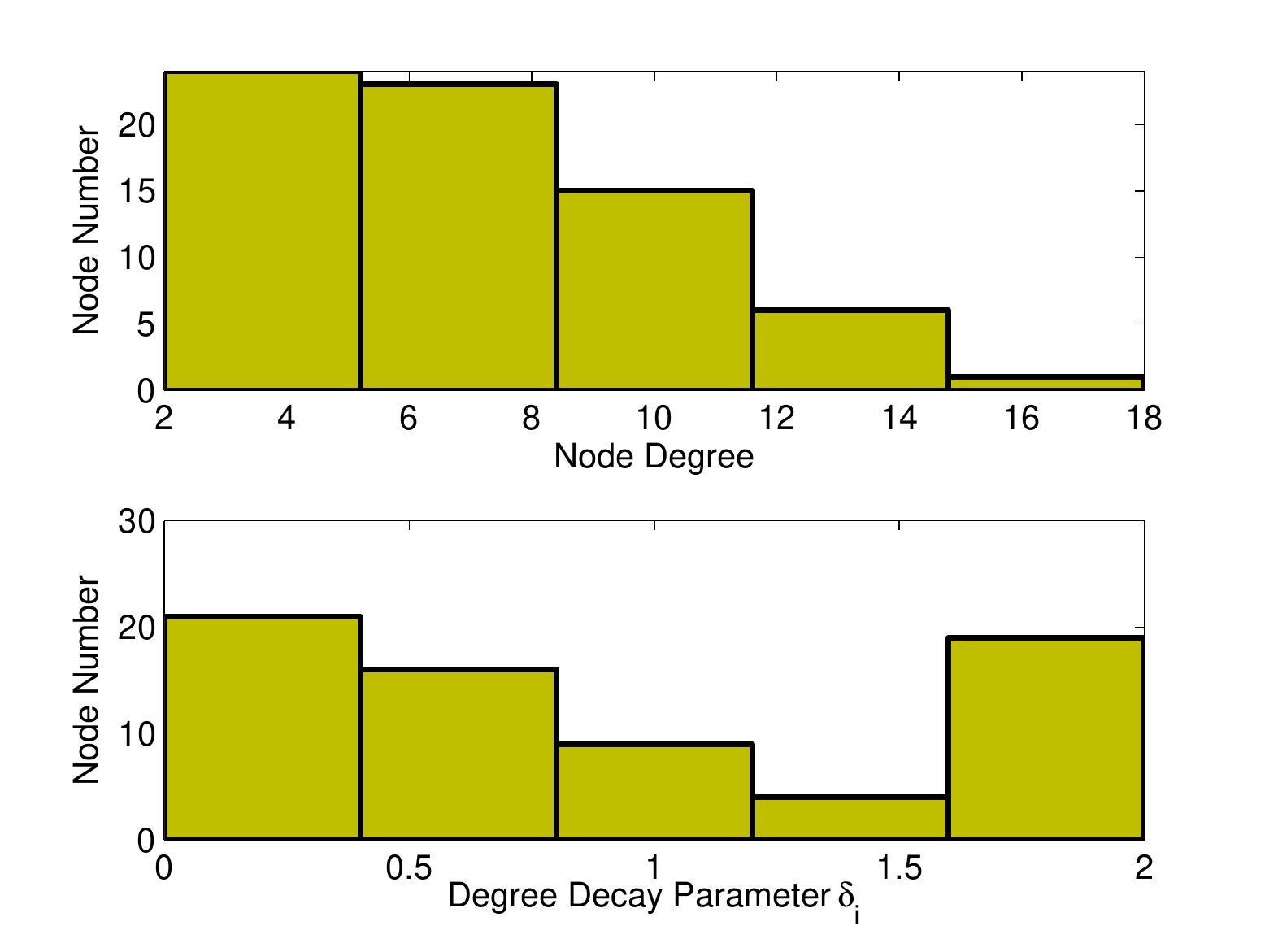}
  \caption{Histograms of Node Degree and Estimated Degree Decay Parameter $\bar \delta_i$ on the Adolescent Health Data.}
  \label{fig:degree_decay_add_health}
\end{figure}

\begin{figure*}[tb]
  \centering
  \begin{subfigure}[b]{0.5\textwidth}
                \centering
                \includegraphics[width=1\textwidth]{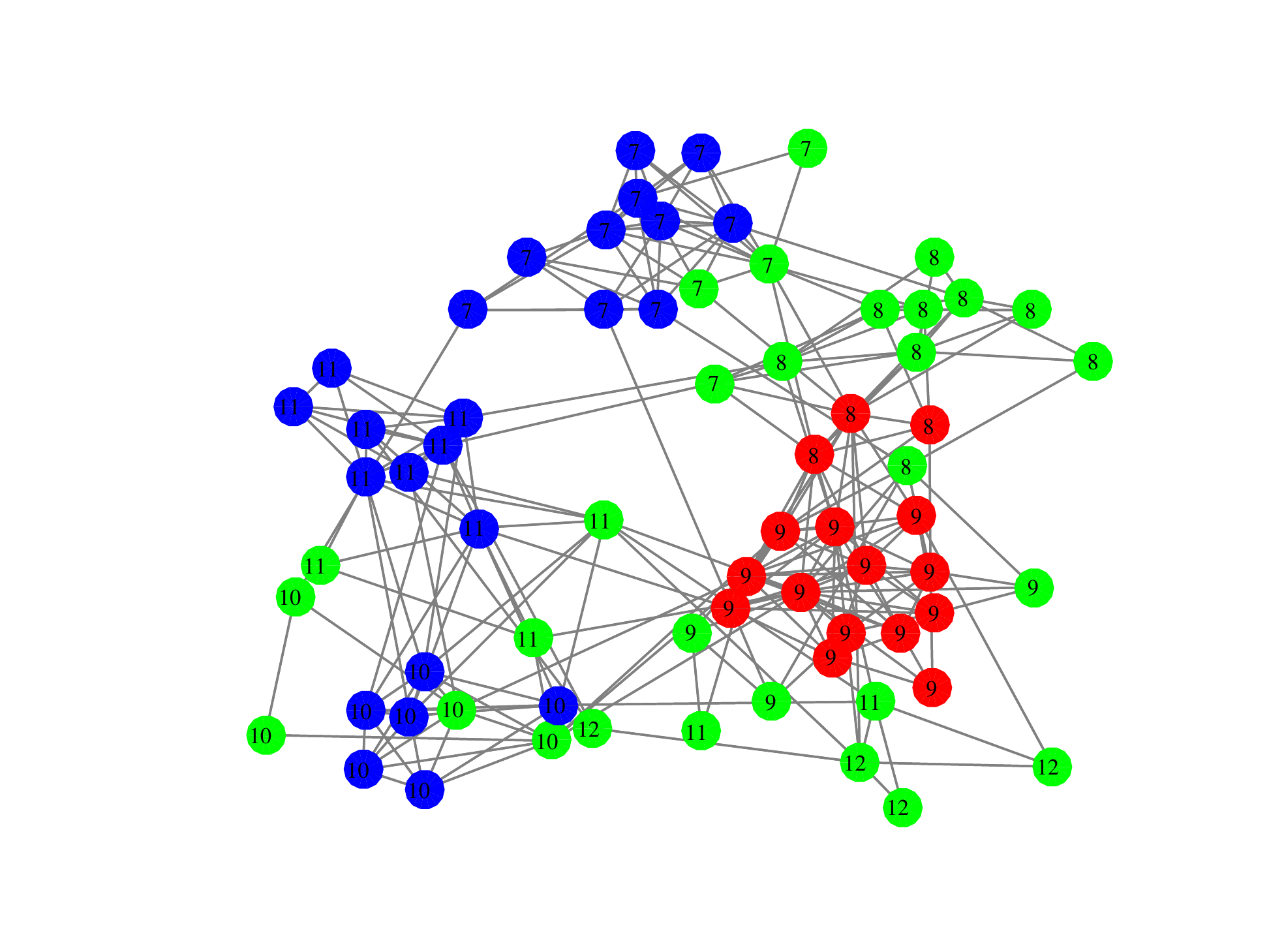}
                \caption{SBM}
                \label{fig:SBM}
  \end{subfigure}
\hspace{-3em}
  \begin{subfigure}[b]{0.5\textwidth}
                \centering
                \includegraphics[width=1\textwidth]{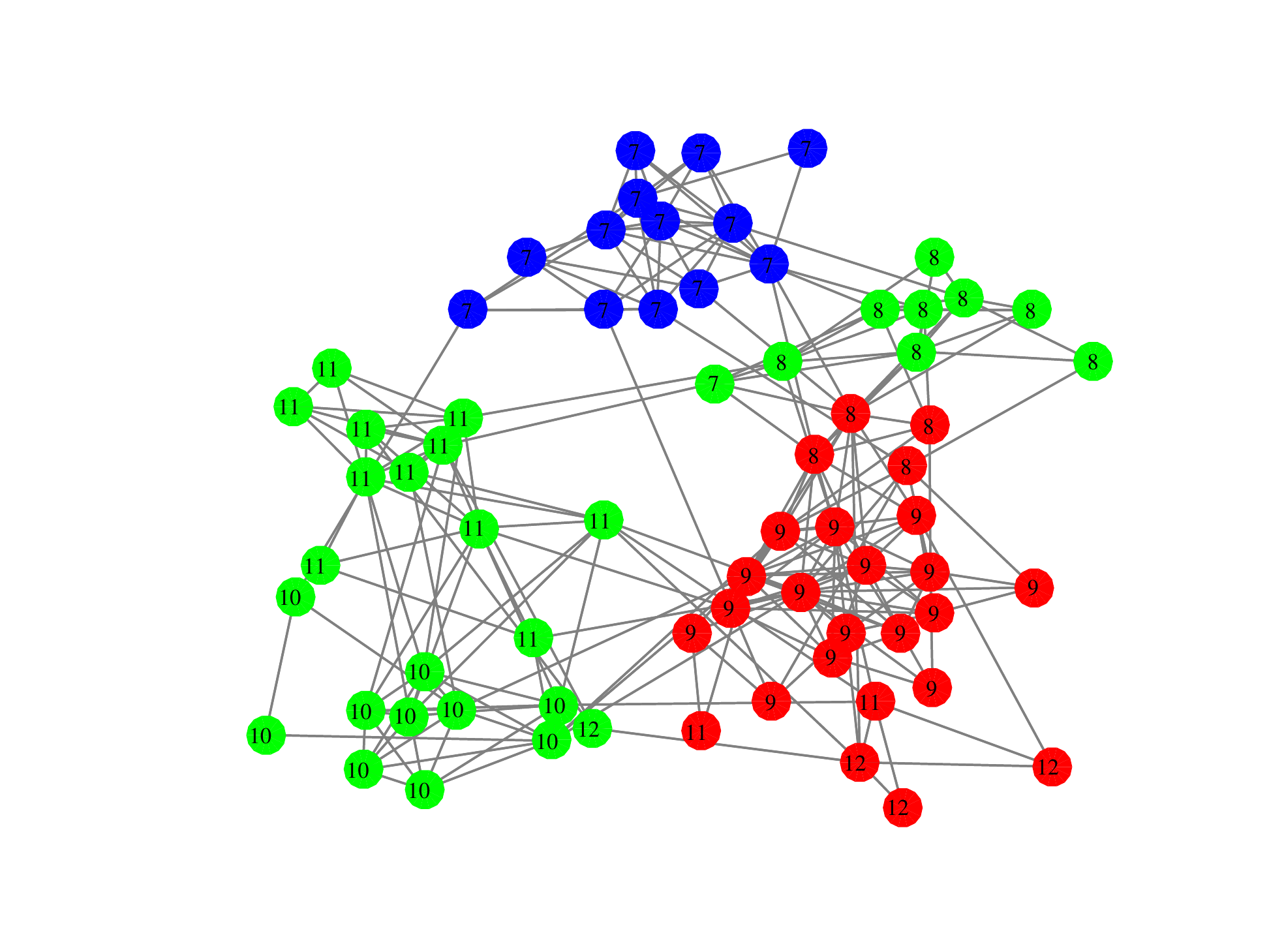}
                \caption{PLD-SBM}
                \label{fig:HD-SBM}
  \end{subfigure}
  \caption{Prediction on the Adolescent Health Data with $K=3$: SBM, stochastic block model \protect\cite{snijders1997estimation}; PLD-SBM, power-law degree stochastic block model. Color stands for predicted cluster; number indicates true grade.}\label{fig:visual_add_health3}
\end{figure*}

First, we implemented the spectral clustering (SC) \cite{ng01spec_clus}\cite{Luxburg:2007TutorialSpeClus} algorithm, based on similarity matrix, in order to see if the generated networks are trivially simple for clustering task.
Its result of clustering error, as shown in Figure \ref{fig:cluster-perf},  indicates that the clustering task is moderately difficult.

Then, both SBM \cite{snijders1997estimation} and two variants, i.e., DC-SBM and PLD-SBM, were also fit to the 20 simulated networks, and the learned models are used to predict the cluster structure. For PLD-SBM, the prediction is obtained by maximizing the variational posterior distribution, i.e.,
  \begin{align} \label{eq:max-cluster}
  z_i^*=\arg\max_{k}\phi_{ik},
  \end{align}
where $\phi_{ik}$ is the optimal variational posterior parameter from (\ref{eq:opt-phi}).
From the clustering error result, as shown in Figure \ref{fig:cluster-perf}, PLD-SBM achieves significant improvement over the three compared models.
We attribute such superiority to the explicitly established power-law representation ability through
the degree decay parameter $\bar \delta_i$.
The histograms of node degrees and the estimated degree decay parameter $\bar \delta_i$ of one simulated network are shown in Figure \ref{fig:degree_decay_simulation}.
shows the histograms of node degree and the estimated degree decay parameter $\bar \delta_i$ on one of the simulated networks. Clearly, the power-law feature of the network is evident, and the degree decay parameter $\bar \delta_i$ varies in a range from $0$ to $1.7$, consistent with the  previous discussion, offering automatic degree adaptation ability, which is absent in SBM.

\section{Real world Application}
\label{sec:adolescent}

\subsection{Adolescent Health Network}

We evaluated PLD-SBM on a real friendship network involving a group of $69$ students in grades $7$ to $12$. The network is drawn from the National Longitudinal Study of Adolescent Health, which is a school-based longitudinal study of the health-related behaviors of adolescents and their outcomes in young adulthood \cite{Harris2003AddHealth}\cite{Udry2003AddHealth}. During the study, an in-school questionnaire was conducted on a sample of students in grades $7$ to $12$ of each school. These students were asked to nominate up to $5$ boys and $5$ girls within the school they regarded as their best friends. The  network we used is from a single school and has been widely used in previous studies \cite{airoldi2008mixed}\cite{handcock2007model}. Note that the original friendship nominations were collected among $71$ students, while $2$ students nominated none. To focus on network connectivity, we simply reformulated the original directed graph, based on friendship nominations, into an undirected setting to train SBM and PLD-SBM.

Firstly, we use the ICL criterion to help choose the number of communities. As shown in Figure \ref{fig:ahd_K}, both baseline model SBM and the proposed model PLD-SBM achieve the highest ICL score with $K=6$, which is equal to the number of grade groups. The node degree characteristic within each grade is obviously skewed, and experiments on such a network demonstrate the effectiveness of the proposed model as shown below.

The degree distribution of the friendship network is plot in Figure \ref{fig:degree_decay_add_health}. It reveals that the network exhibits a significant power-law feature. We therefore expect PLD-SBM to outperform SBM in identifying cluster structures hidden in the network, by addressing the skew degree distribution within each cluster. Like previous studies, we used grade as the true cluster index. We also implemented two more SBM variants, i.e., the mixed membership stochastic model (MMSBM) \cite{airoldi2008mixed} and the stochastic block mixture model (SBMM) \cite{Doreian07Discussion}, for comparison. The prediction results are presented in Table \ref{tab:cluster-perf}. PLD-SBM outperforms SBM by reducing the 13 miss-predicted errors to 7, and one clustering result is visualized via the Fruchterman-Reingold algorithm \cite{salter2012review} and shown in Figure \ref{fig:visual_add_health}. We attribute this superiority to its ability of power law modeling, and the fitted degree decay parameters $\bar \delta_i$ vary considerably, as shown in Figure \ref{fig:degree_decay_add_health}. Similarly, PLD-SBM also outperforms SBMM in terms of clustering error as shown in Table \ref{tab:cluster-perf}. However, its performance is slightly inferior to MMSBM. This might due to some mixed membership structures indeed existing in this network.

In addition, we set $K=3$ to further examine the clustering behavior of baseline SBM and the proposed PLD-SBM. The cluster structures discovered by the two are shown in Figure \ref{fig:SBM} and \ref{fig:HD-SBM} respectively. PLD-SBM consistently retains the skewness of degree distribution throughout the network. By contrast, SBM divides the network into groups of high-degree `hub' nodes and of low-degree `peripheral' nodes. Such division is obviously not suitable here.

\subsection{the Political Blog Network}
\label{sec:other}

We also evaluated PLD-SBM on a larger real-world network, political blogs, constructed by Adamic and Glance \cite{adamic2005political}. The nodes, from the front pages of individual and group blogs were labeled as either liberal or conservative according to the blog's political leanings. The edges were retrieved as URL references. Like \cite{karrer2011stochastic}, we treated the directed-constructed network as an undirected form and considered only the largest connected subgraph, assembled by $1,222$ nodes and $19,089$ edges.

\begin{figure}[tb]
    \centering
    \includegraphics[width=1\columnwidth]{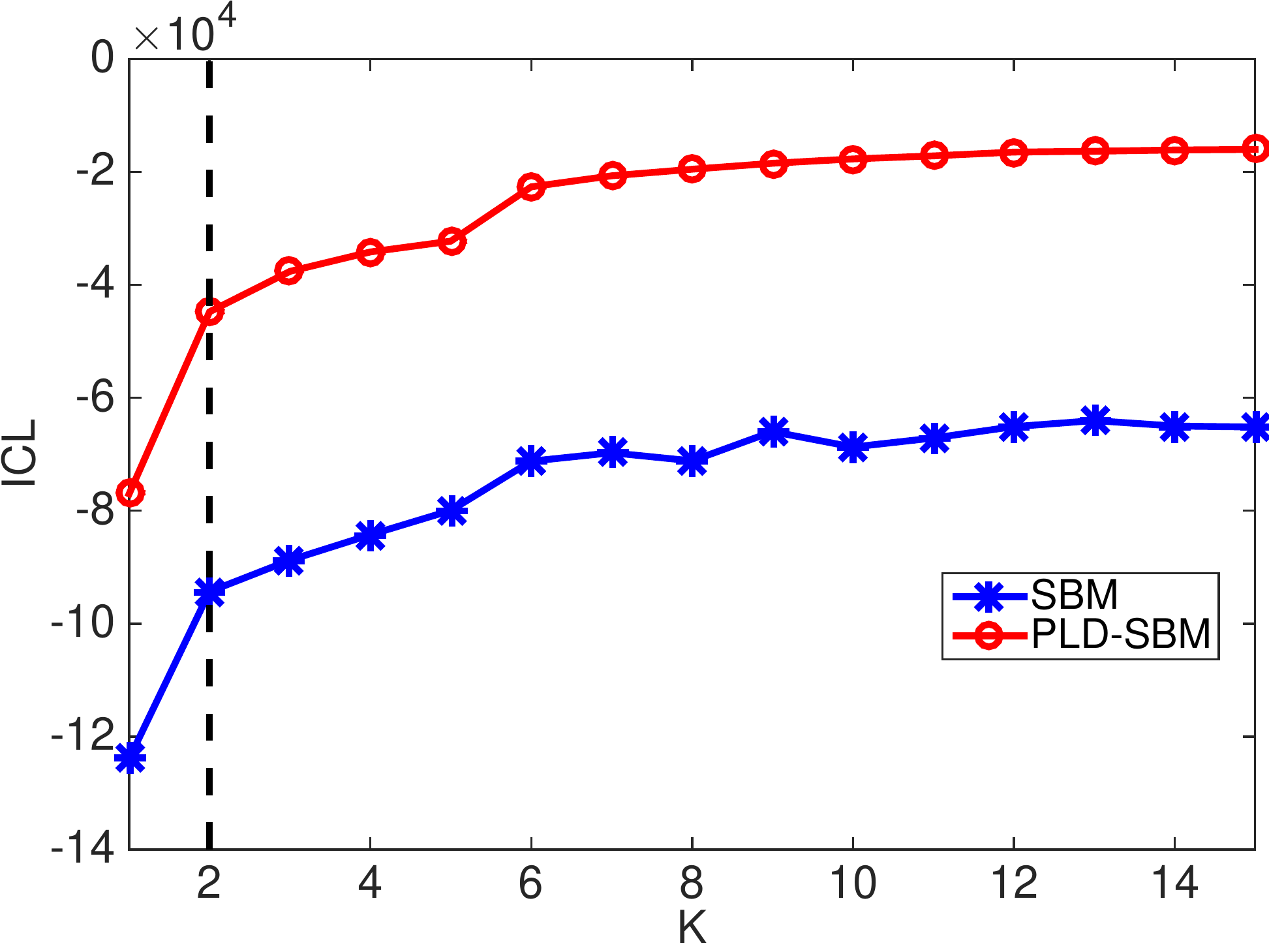}
    \caption{Model selection of $K$ on the political blog network. Although $K=2$ does not achieve the best ICL scores for both SBM and PLD-SBM, it improves $K=1$ greatly. Here, due to the larger size of this dataset (comparing to the Adolescent Health Data), it is natural to use more complex model to fit. This is the reason for the upward trending of both curves. }\label{fig:pbn_K}
\end{figure}

The ICL scores with different specifications of $K$ for this dataset is shown in Figure \ref{fig:pbn_K}. The curves clearly show that the proposed PLD-SBM consistently outperforms the baseline SBM in terms of likelihood with varying $K$s. In addition, the upward trending indicates that the ICL score prefers larger $K$ for this dataset. However, to dig deeper how PLD-SBM works differently from the state-of-the-art models, we choose smaller values, i.e., $K=2,3,4,5$, to visually demonstrate the different community structures discovered by different models.

\begin{figure}[tb]
  \centering
  \includegraphics[width=1\columnwidth]{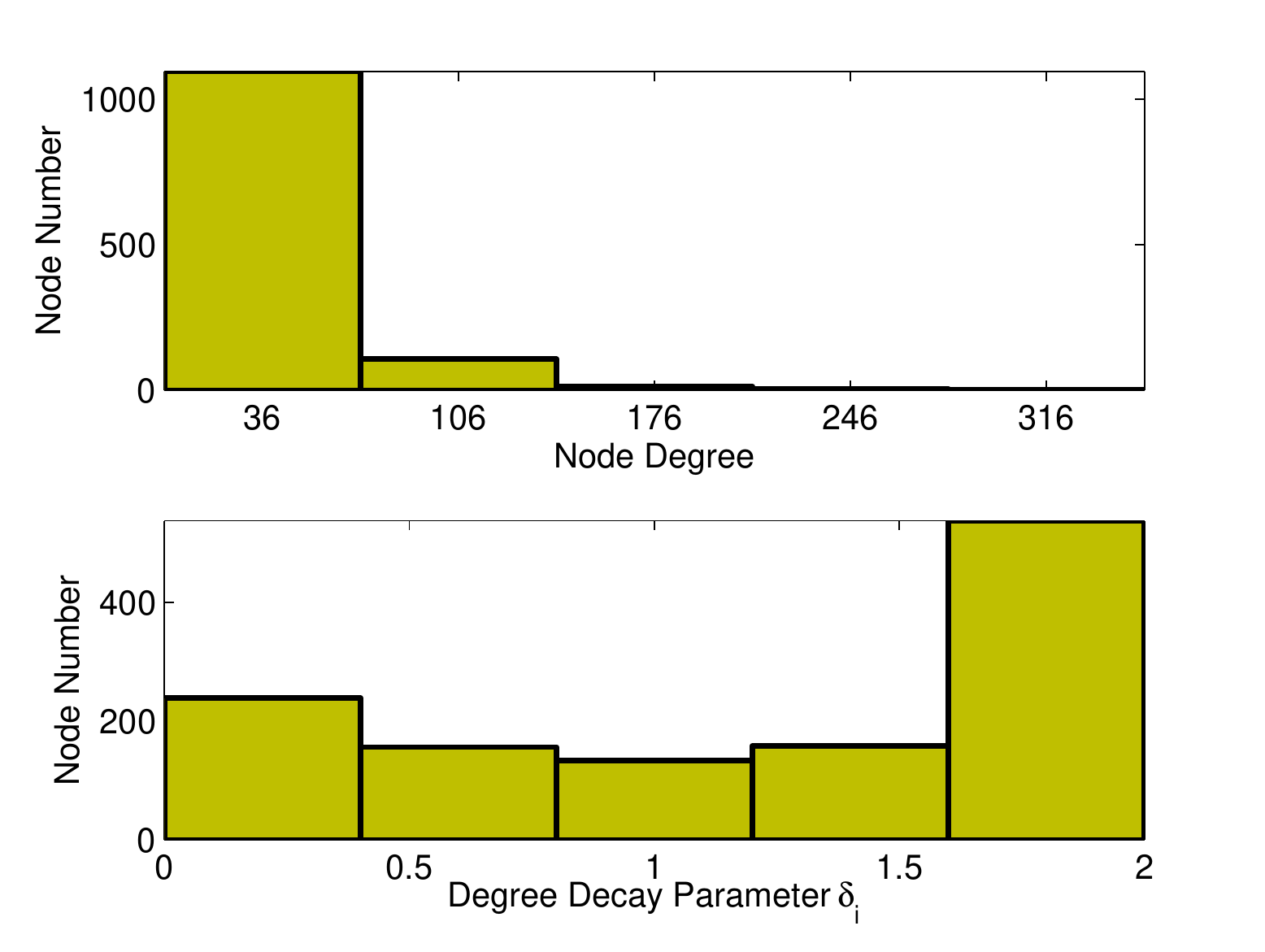}
  \caption{Histograms of Node Degree and Estimated Degree Decay Parameter $\bar \delta_i$ on the political blog network.}\label{fig:degree_decay_polblog}
\end{figure}

Firstly, we evaluate the case $K=2$, which is equal to the true number of political parties in the network. The degree distributions for both political parties are highly skewed, as demonstrated in Figure \ref{fig:degree_decay_polblog} by the histogram of the overall degree distribution. This is consistent with the intuition that each political party has only a few popular blogs of over a hundred links and the rest of the blogs having rare connections with other blogs. We therefore expect that PLD-SBM will fit this network better than SBM. This is verified by their clustering accuracies reported in Table \ref{tab:cluster-perf-polblog}. PLD-SBM achieves a clustering error of $0.0466$, while the baseline SBM outputs higher error rate of $0.4492$. Again, we attribute this superiority to its ability of addressing the power-law feature, and the range of fitted degree decay parameters $\bar \delta_i$ varies considerably, shown in Figure \ref{fig:degree_decay_polblog}, from $0$ to $2$.

\begin{table}
  \centering
  \renewcommand{\arraystretch}{1.8}
  \setlength{\tabcolsep}{12pt}
  \begin{tabular}{r|*{2}{c}|*{2}{c}}
  &\multicolumn{2}{|c|}{SBM } &\multicolumn{2}{|c}{PLD-SBM }  \\
  Grade&1&2&1&2\\
    \hline
    Conservative   &329&258    &548&39    \\
    Liberal   &291&344    &18&617   \\
    \hline
    &\multicolumn{2}{|r|}{Error Rate 0.4492}&\multicolumn{2}{|r|}{Error Rate 0.0466}
  \end{tabular}
  \caption{Performance Evaluation on the Political Blog Network: SBM, stochastic block model \protect\cite{snijders1997estimation}; PLD-SBM, power-law degree stochastic block model.}\label{tab:cluster-perf-polblog}
\end{table}

\begin{figure*}[tb]
  \centering
  \begin{subfigure}[b]{0.24\textwidth}
                \centering
                \includegraphics[width=1\textwidth,angle = 15]{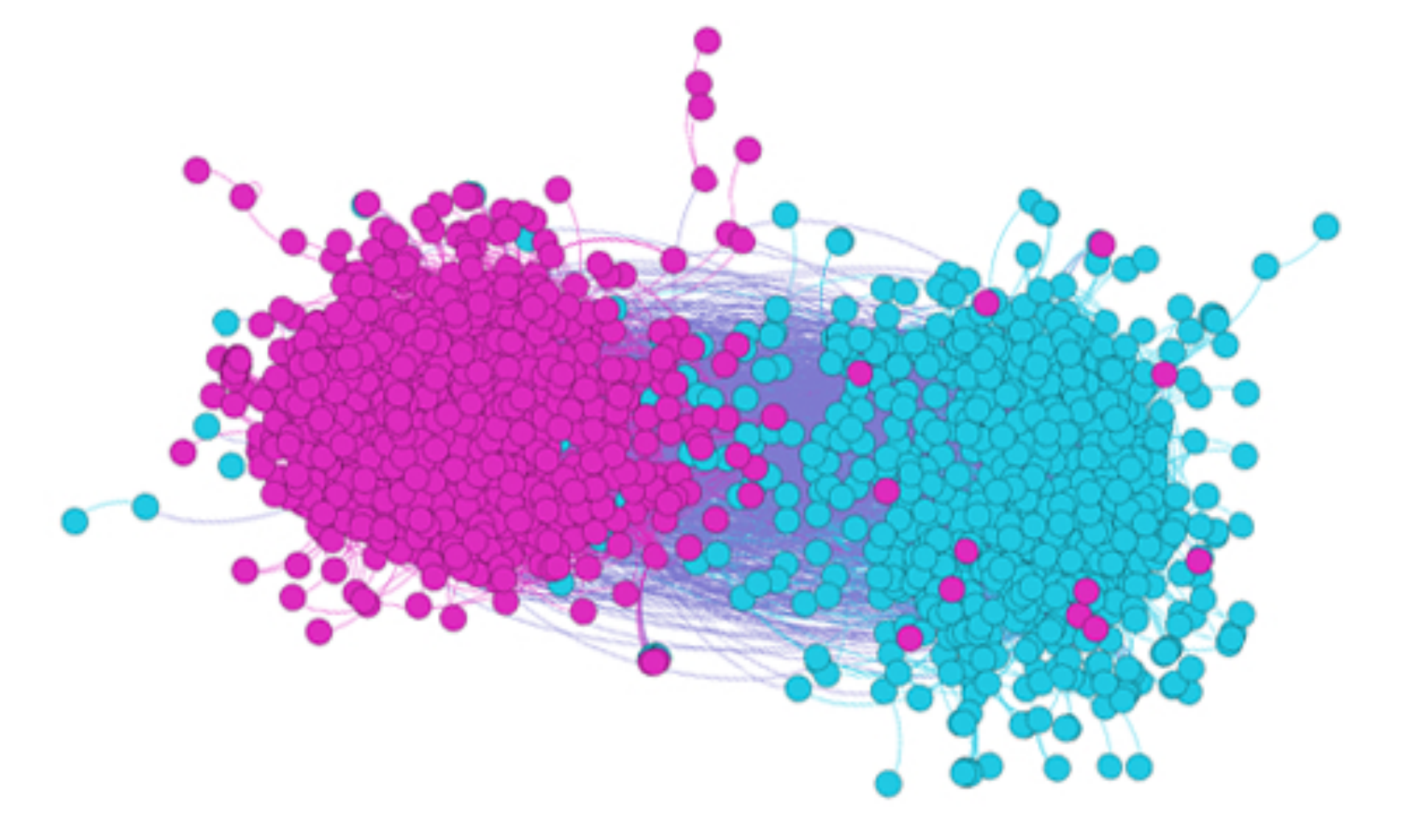}
                \caption{Truth}
                \label{fig:truth-polblog}
  \end{subfigure}%
  \begin{subfigure}[b]{0.24\textwidth}
                \centering
                \includegraphics[width=1\textwidth,angle = 15]{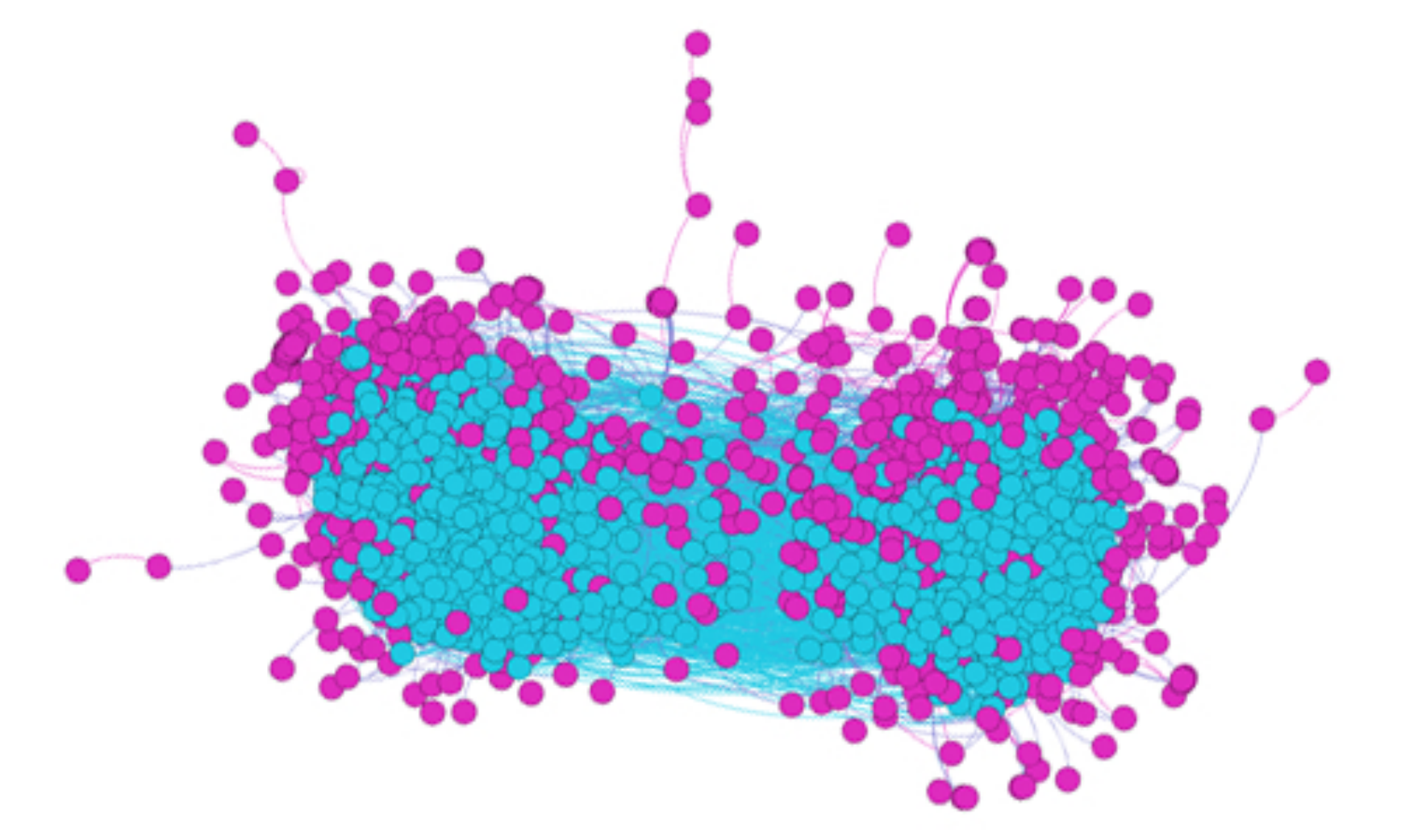}
                \caption{SBM}
                \label{fig:SBM-polblog}
  \end{subfigure}%
   \begin{subfigure}[b]{0.24\textwidth}
                \centering
                \includegraphics[width=1\textwidth,angle = 15]{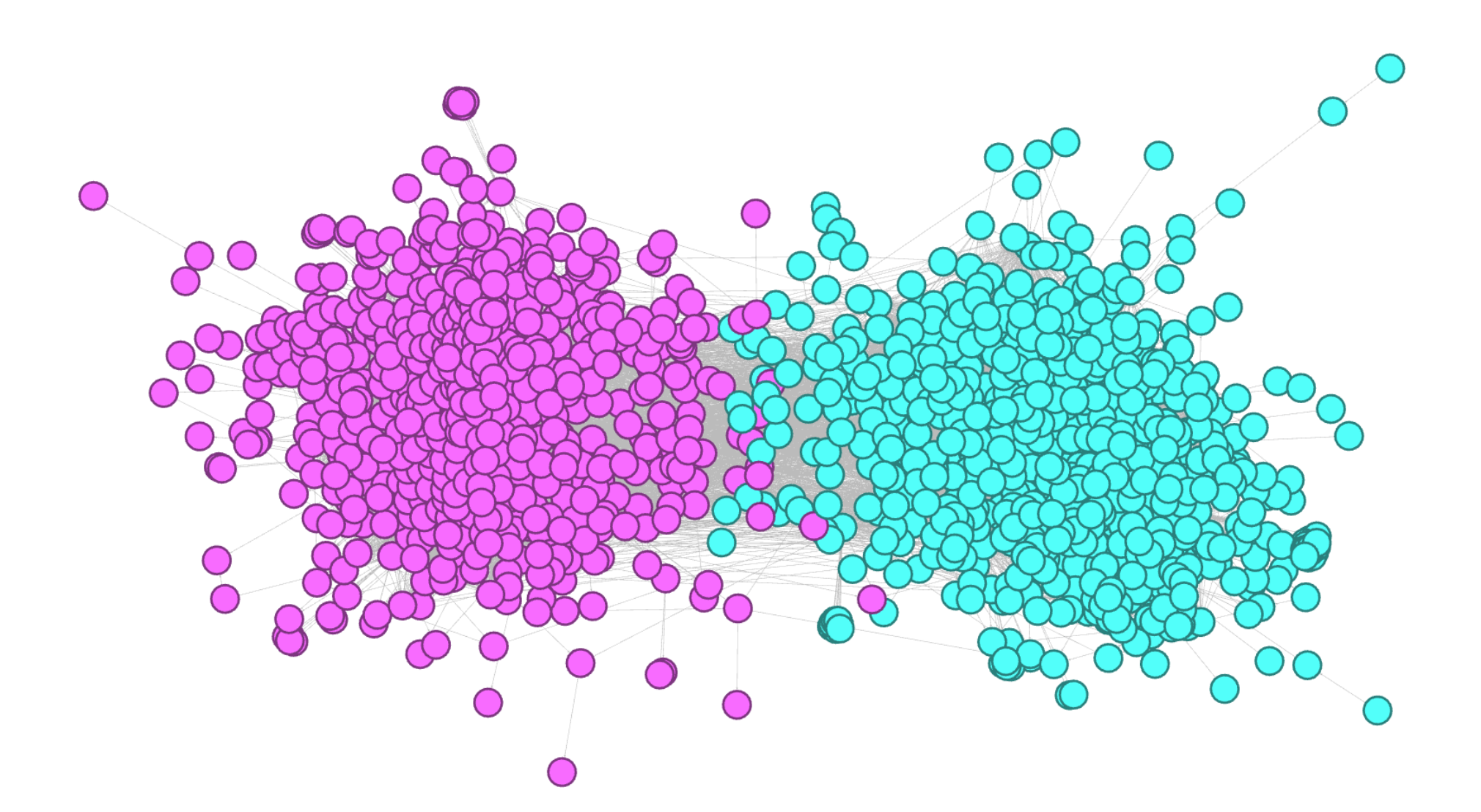}
                \caption{DC-SBM}
                \label{fig:DCSBM-polblog}
  \end{subfigure}%
  \begin{subfigure}[b]{0.24\textwidth}
                \centering
                \includegraphics[width=1\textwidth,angle = 15]{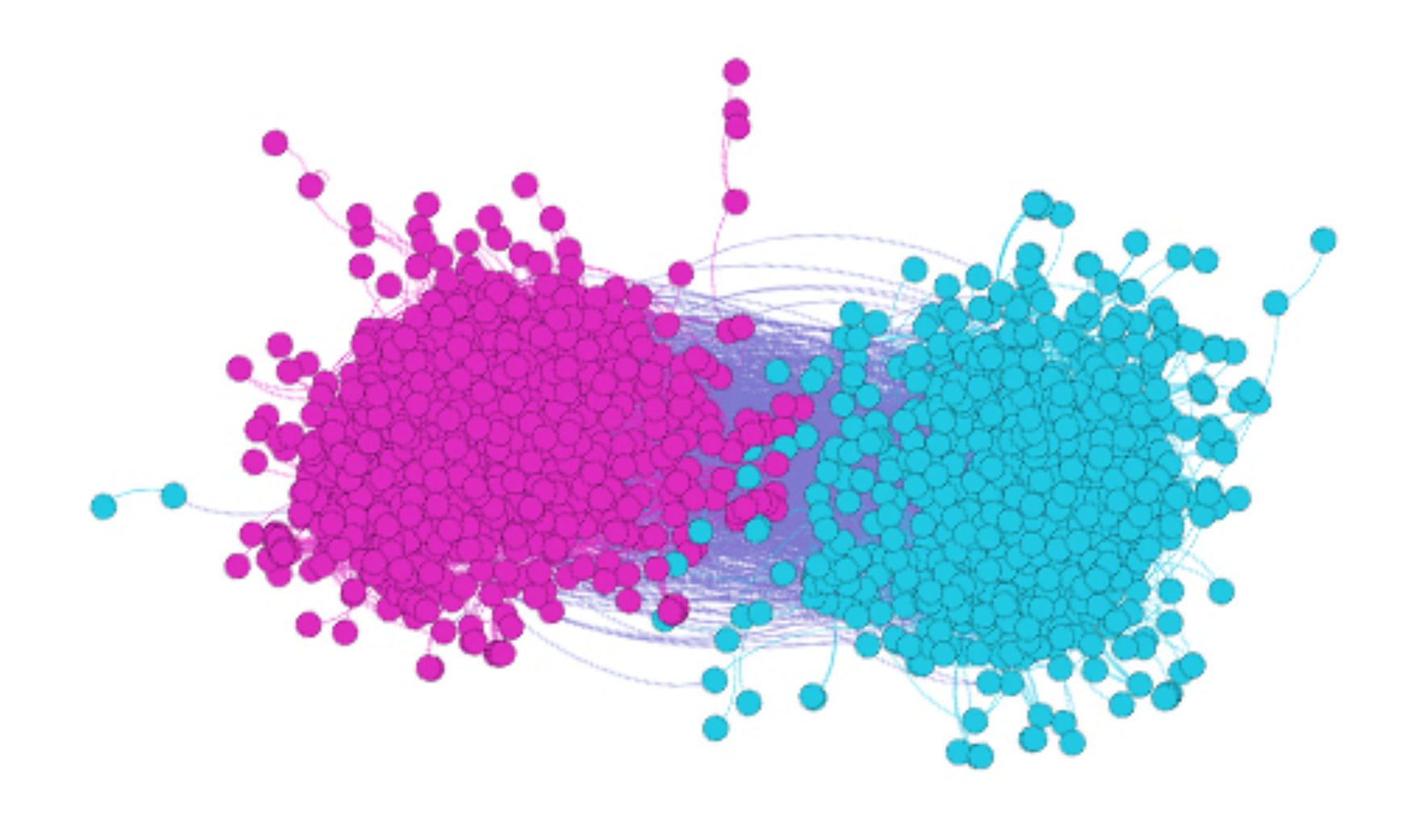}
                \caption{PLD-SBM}
                \label{fig:PLDSBM-polblog}
  \end{subfigure}
  \caption[a]{Prediction on the Political Blog Network: Truth, manually labeled by \cite{adamic2005political};
   SBM, stochastic block model \cite{snijders1997estimation};
   DC-SBM, degree-corrected stochastic block model \cite{karrer2011stochastic};
 PLD-SBM, power-law degree stochastic block model.
 Cyan represents liberal blog nodes, and purple represents conservative blog nodes.
 }
\label{fig:visual_polblog}
\end{figure*}

Figure \ref{fig:visual_polblog} visualizes the cluster structures by ForceAtlas graph layout algorithm \cite{jacomy2014forceatlas2}. Observe that SBM obtained low- and high- degree node groups
indicating the SBM was unable to recover the hidden cluster structures. On the contrary, PLD-SBM inferred a better cluster structure and it was very close to the ground-truth manually labeled by Adamic and Glance \cite{adamic2005political}. We also compared our proposed PLD-SBM with a degree-corrected extension of SBM, i.e., DC-SBM. PLD-SBM achieved a normalized mutual information (NMI) \cite{white2004performance} of $0.7323$ while DC-SBM achieved a value of $0.72$ (reported in \cite{karrer2011stochastic}). According to this, PLD-SBM outperformed DC-SBM. We attribute this slight superiority of PLD-SBM to its devotion to dealing with the power-law distribution, which coincides fairly well with the degree distribution in the political blog network. DC-SBM, on the other hand, is supposed to handle much more general degree heterogeneity and therefore loses its advantage in this case.

\begin{figure}[h]
\begin{center}$
\begin{array}{ccc}
 K=3 & K=4 &K=5
\\
\includegraphics[width=25mm]{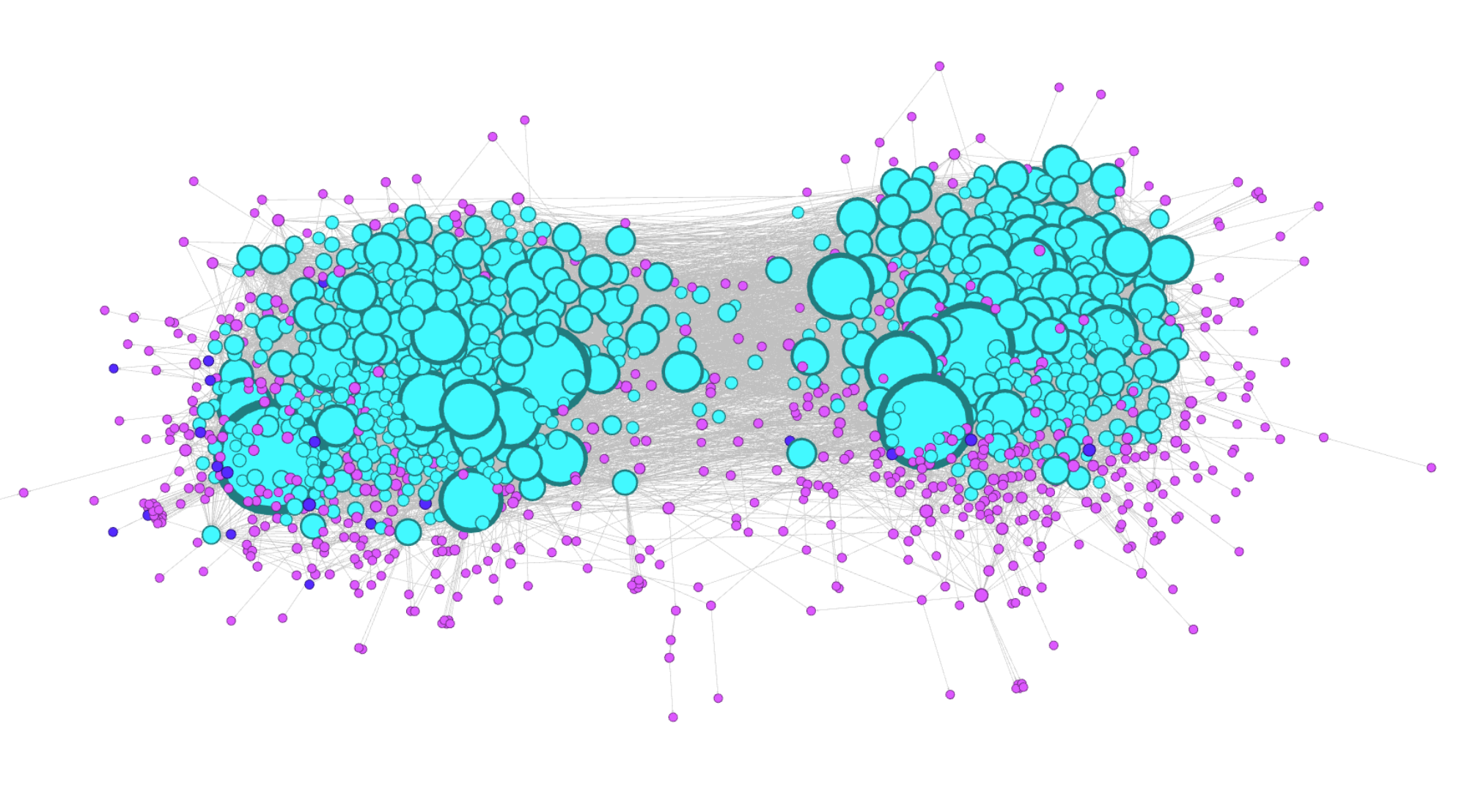}&
\includegraphics[width=25mm]{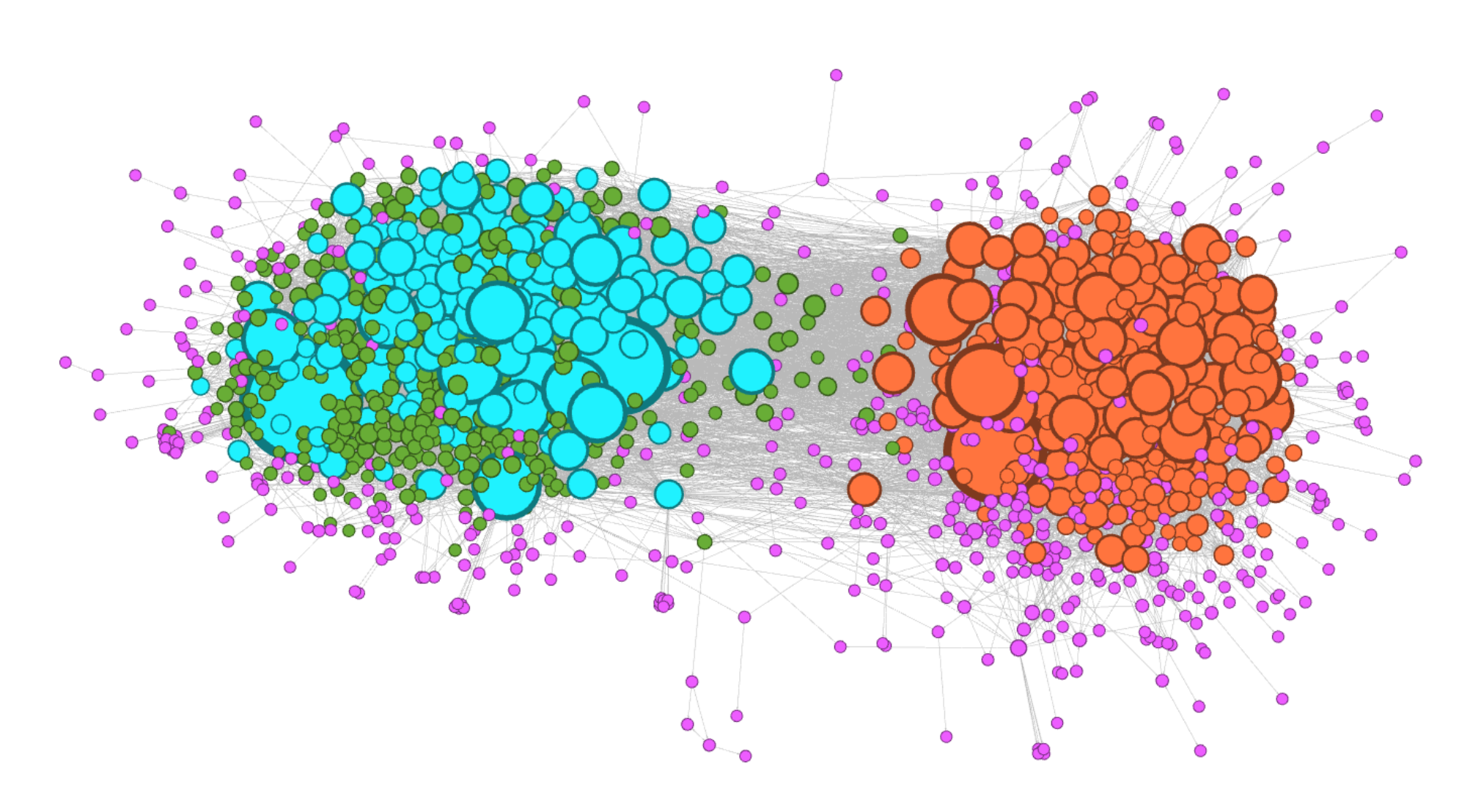}&
\includegraphics[width=25mm]{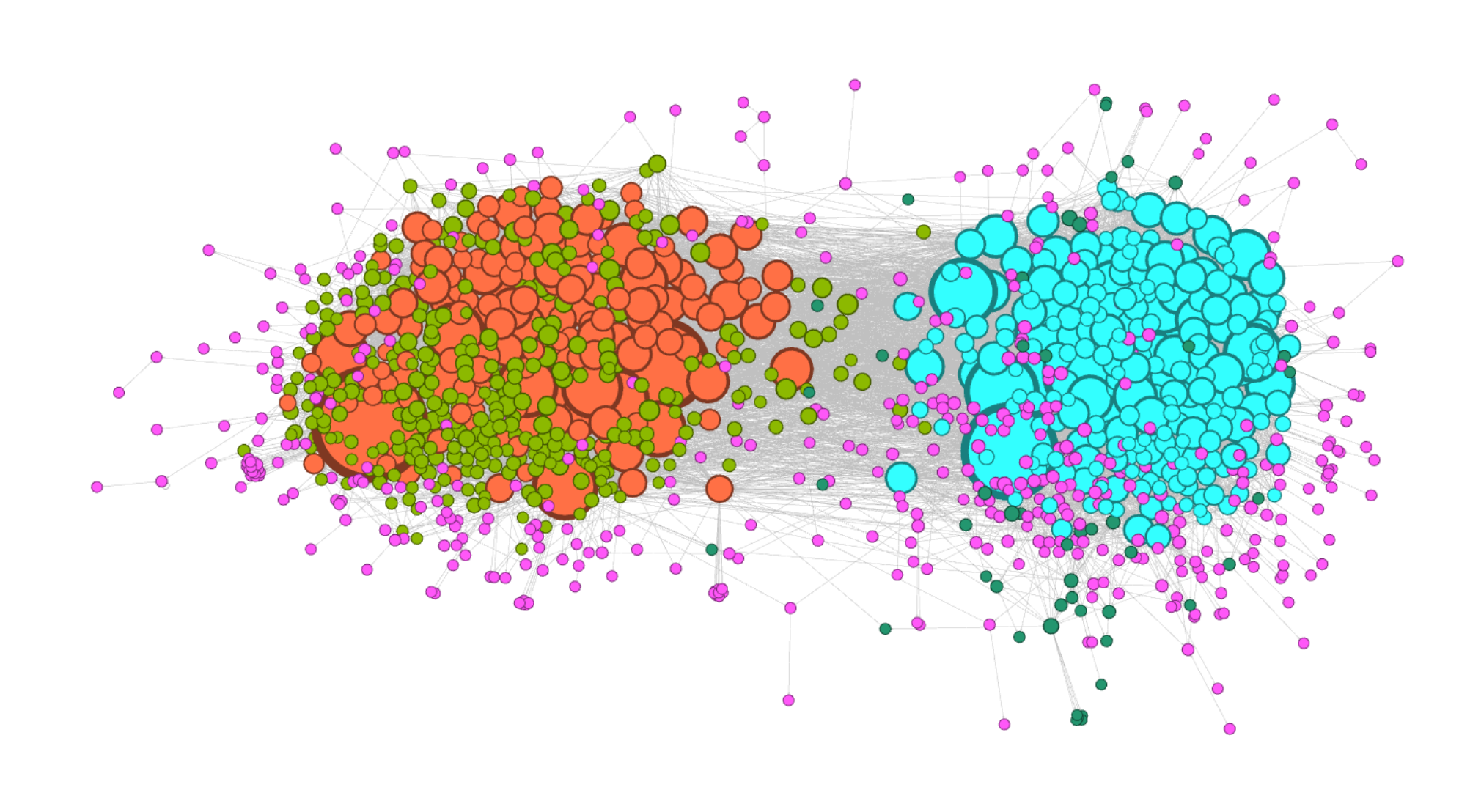}
\\
\includegraphics[width=25mm]{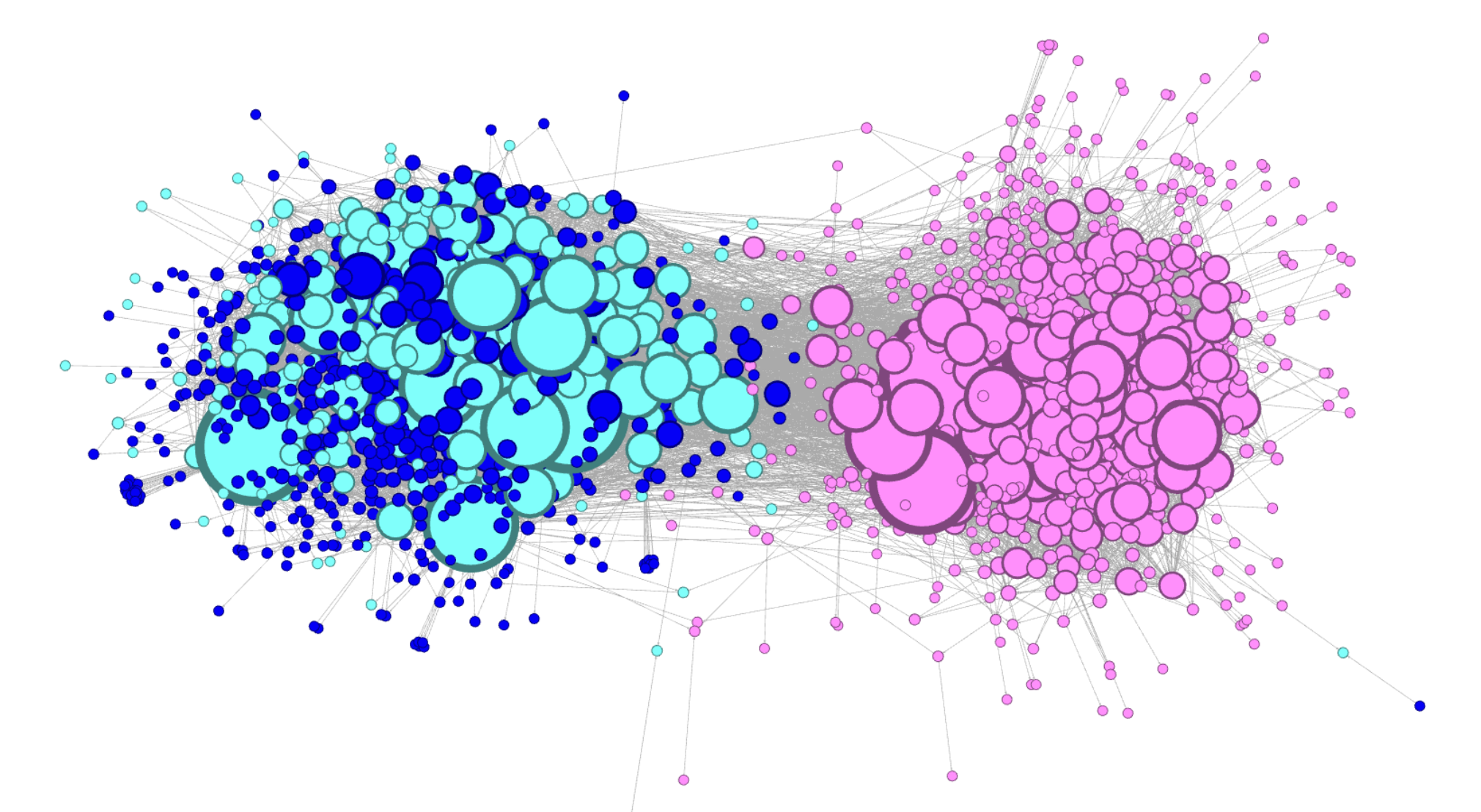}&
\includegraphics[width=25mm]{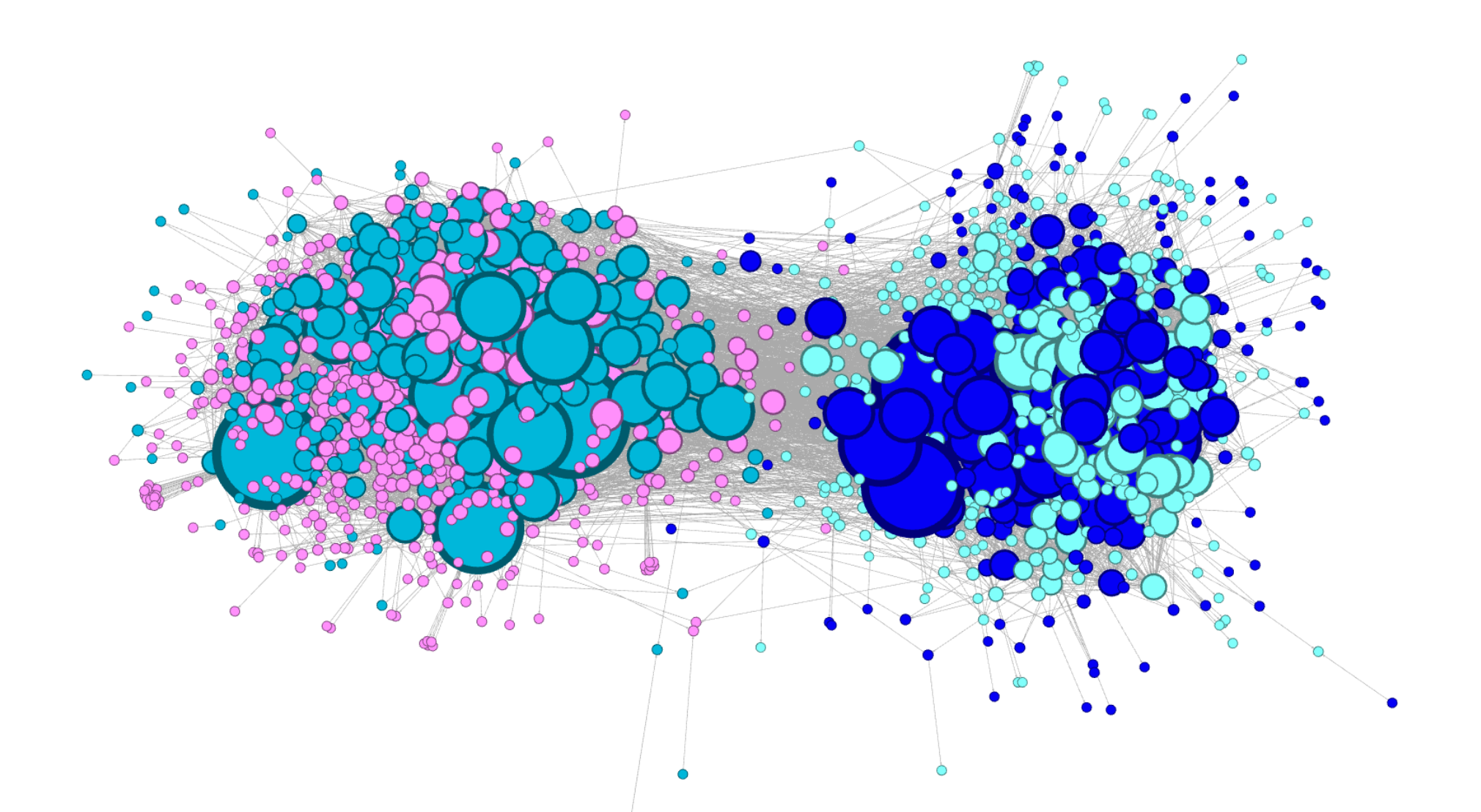}&
\includegraphics[width=25mm]{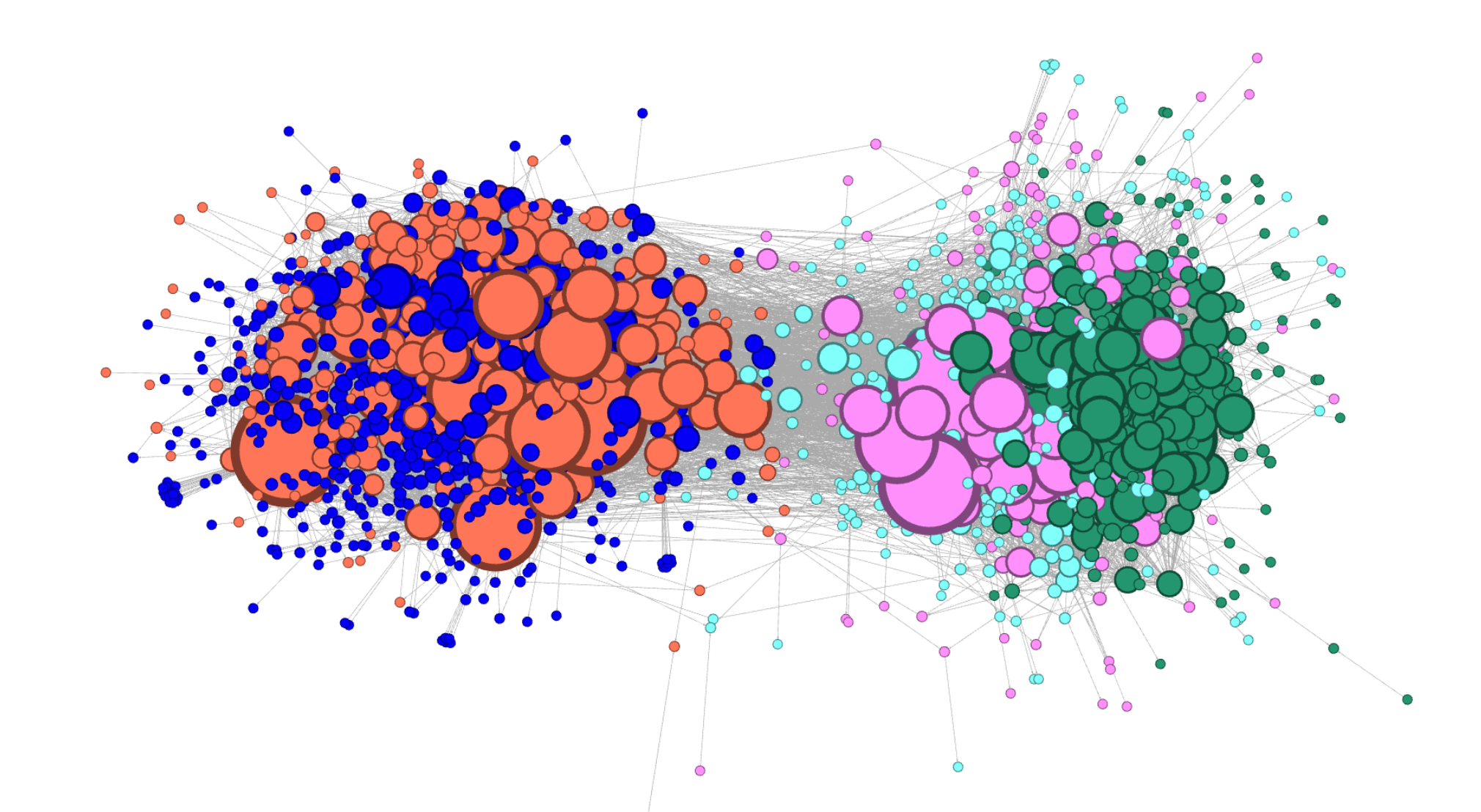}
\\
\includegraphics[width=25mm]{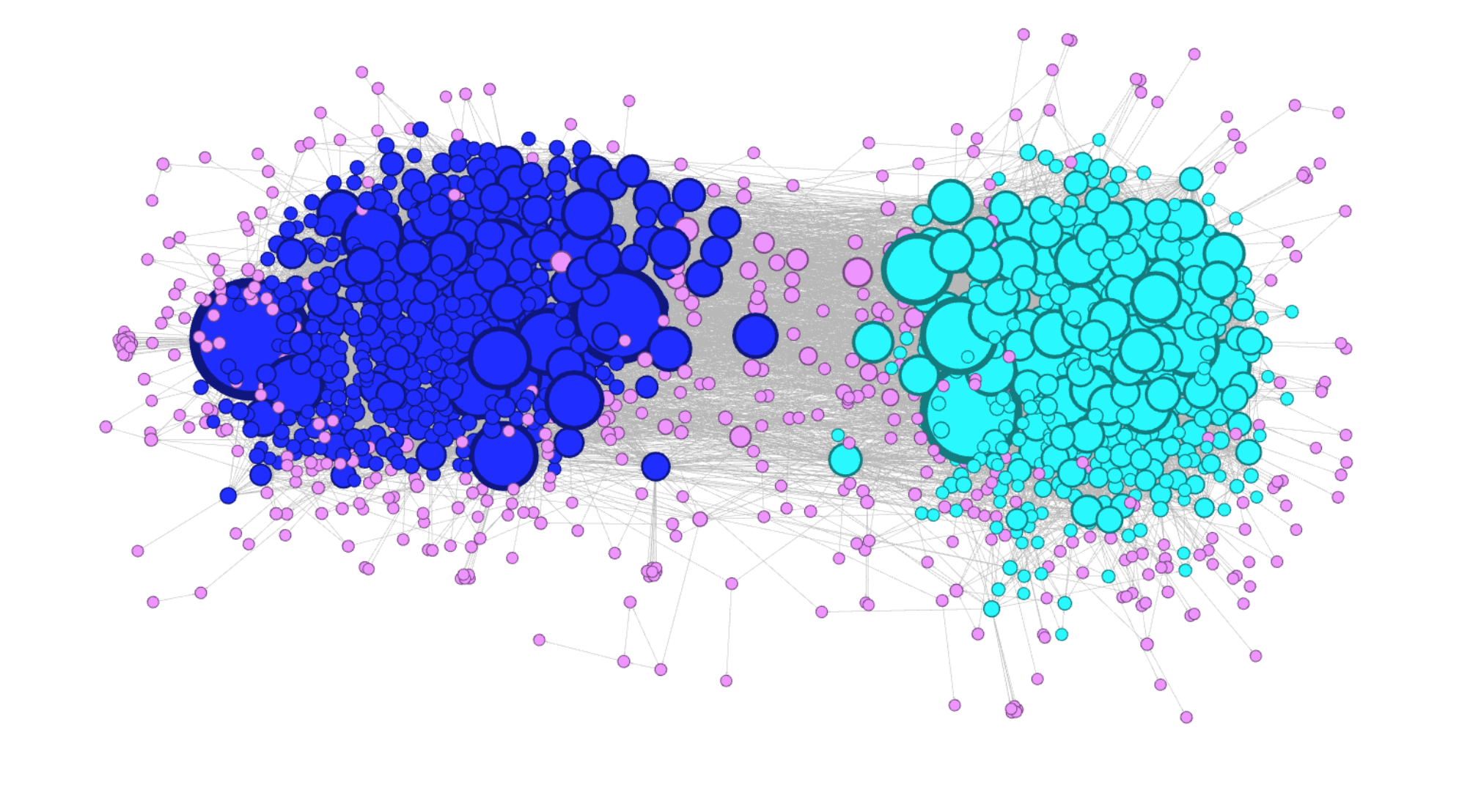}&
\includegraphics[width=25mm]{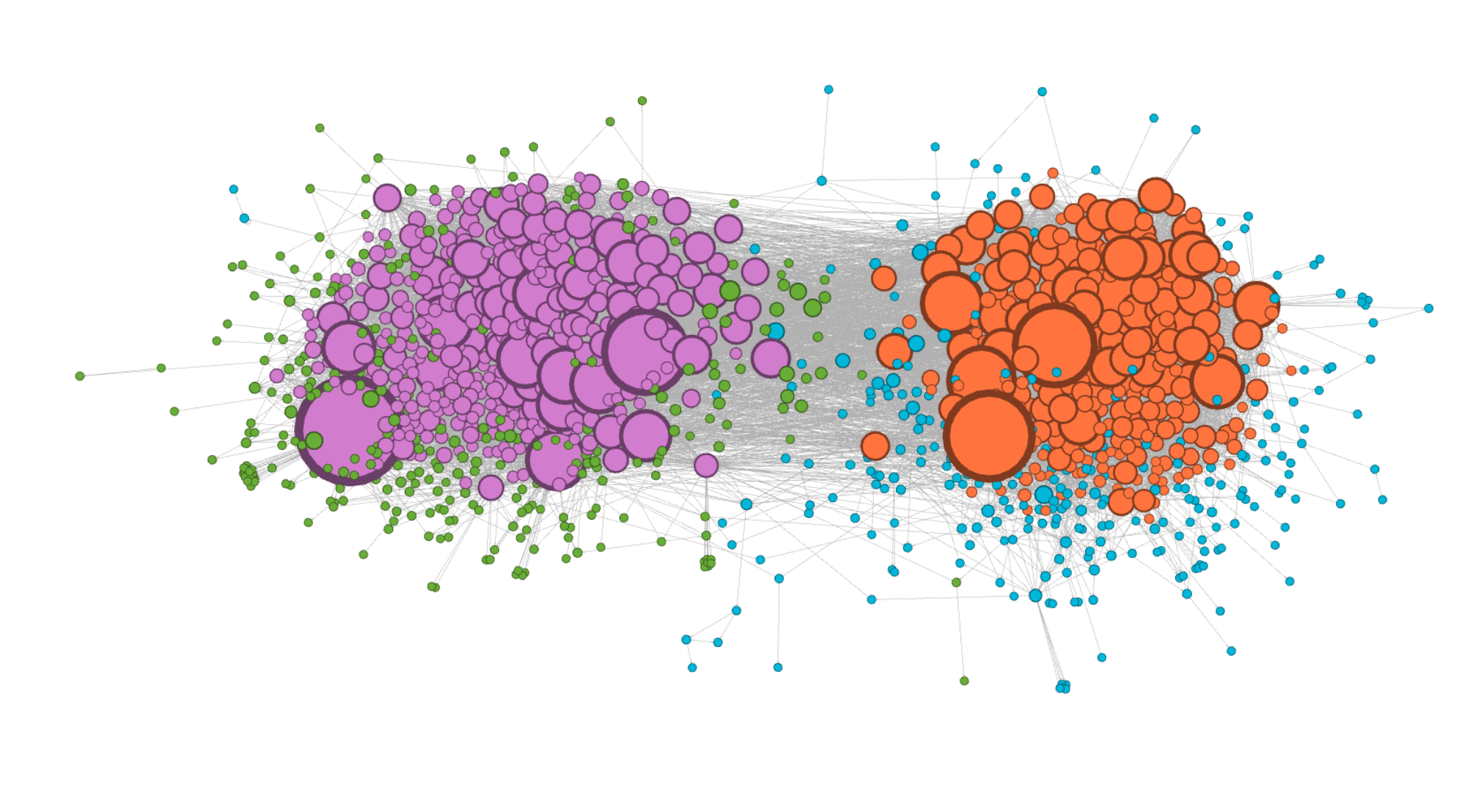}&
\includegraphics[width=25mm]{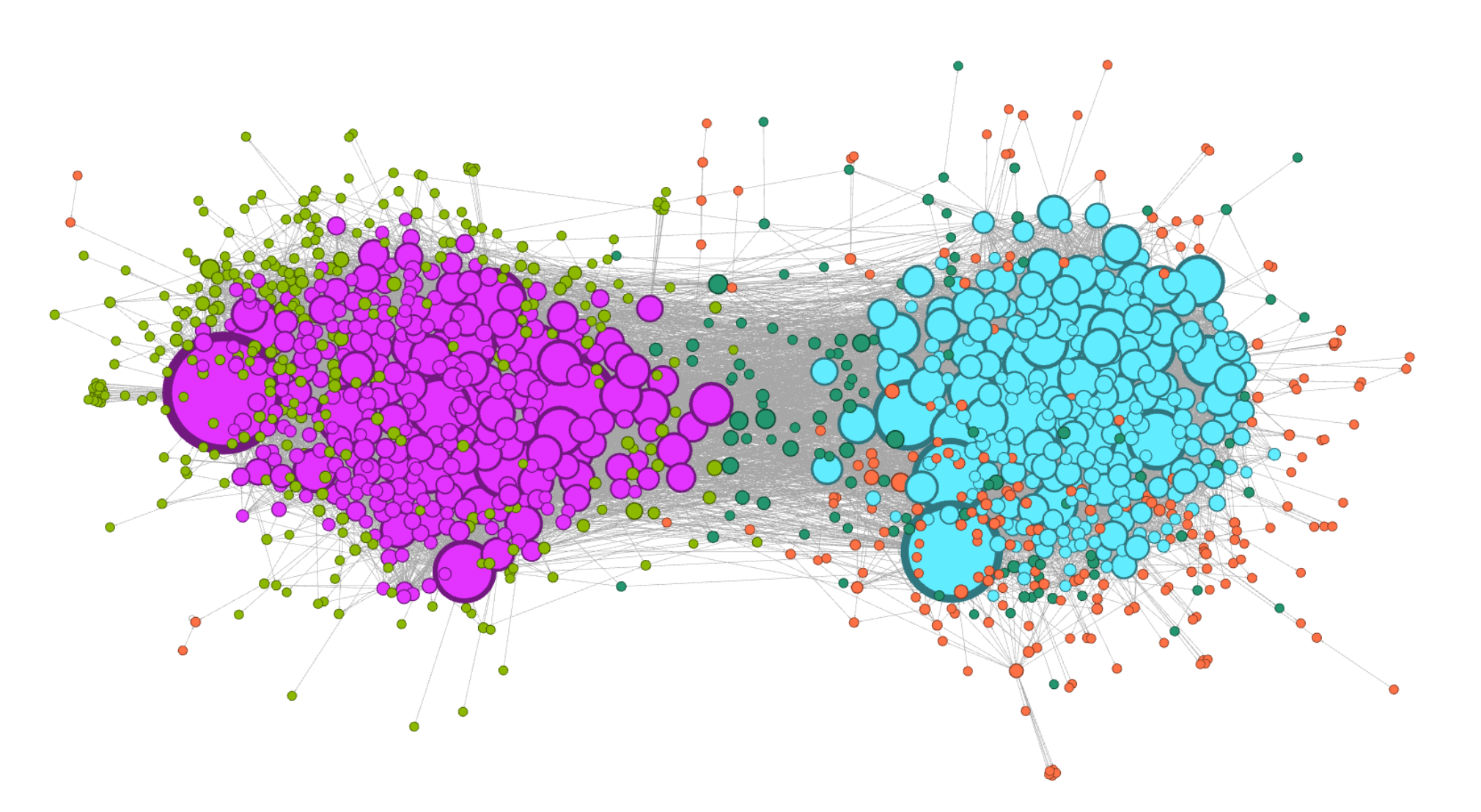}
\\
\end{array}$
\end{center}
\caption{Clustering results on political blog network with different Ks: $1$st row for SBM, $2$nd row for DC-SBM, $3$rd row for PLD-SBM.}
\label{fig:politicalwithvaryingK}
\end{figure}

Finally, we compare SBM, DC-SBM and PLD-SBM in terms of cluster structures with $K=3,4,5$ to explore more insight about their different working mechanism. The results are visualized in Figure \ref{fig:politicalwithvaryingK}.
SBM behaviours consistently throughout different $K$s. Nodes with similar degrees
are grouped together. As $K$ increases, the low-degree peripheral nodes coloured in purple, which are more than high-degree ones, are further divided into smaller groups. To address the homogeneity issue of cluster node degrees inherited in SBM, both DC-SBM and PLD-SBM are proposed but they adopt different strategies as discussed previously. Thus, the cluster structures discovered by them, when $K$ varies from $3$ to $5$, should be different. In the second row in Figure \ref{fig:politicalwithvaryingK}, DC-SBM divides large groups with heterogenous node degrees into smaller ones with consistent heterogenous node degrees, as $K$ increases. In comparison, PLD-SBM retains the two political parties, but splits the peripheral nodes into smaller clusters as $K$ increases, which are demonstrated in the third row of the figure.

\section{Conclusion}
\label{sec:conclusion}

A new extension of stochastic block models (SBM), termed power-law degree SBM (PLD-SBM), has been developed in this paper. By adding a layer of hidden variables associated with degree decay of every single node, the proposed model exhibits power-law degree characteristic and explicitly addresses the ubiquitous scale-free feature of real networks. Such a property enables PLD-SBM to correct the homogeneous degree bias of SBM. Experiments conducted on both simulated networks and real-world networks, i.e., a friendship network from the National Adolescent Health data and the political blog network, verity the effectiveness of the proposed PLD-SBM.

\section*{Acknowledgment}
This work was supported in part by the National Natural Science Foundation of China under Grants
61702145, 61622205, and 61472110, the Zhejiang Provincial Natural Science Foundation of China under Grant LR15F020002, the Australian Research Council Projects under Grant FL-170100117, DP-180103424, DP-140102164, and LP-150100671.

\ifCLASSOPTIONcaptionsoff
  \newpage
\fi

\bibliographystyle{IEEEtran}
\bibliography{bib}

\begin{IEEEbiography}[{\includegraphics[width=1in,height=1.25in,clip,keepaspectratio]{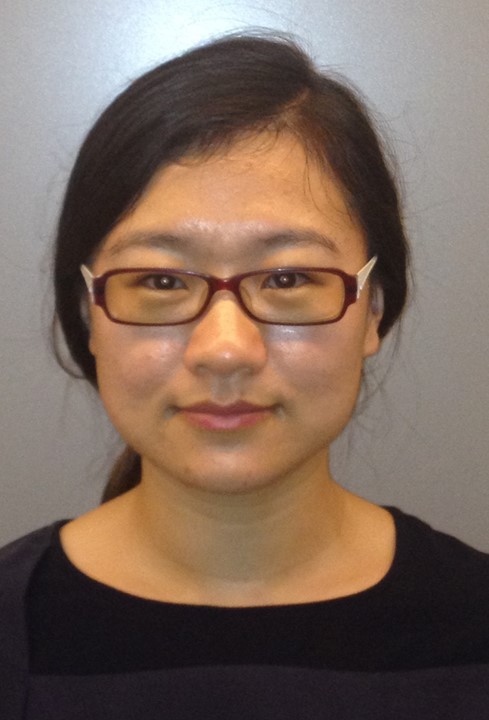}}]
{Maoying Qiao}
received the B.Eng. degree in Information Science and Engineering from Central South University, Changsha, China, in 2009, and the M.Eng. degree in Computer Science from Shenzhen Institutes of Advanced Technology, Chinese Academy of Sciences, Shenzhen, China, in 2012, and the PhD degree in Computer Science in 2016 fro the University of Technology Sydney. She is currently an Associate Professor with the School of Computer Science and Technology, Hangzhou Dianzi University.
Her research interests include machine learning and probabilistic graphical modeling.
\end{IEEEbiography}

\begin{IEEEbiography}[{\includegraphics[width=1in,height=1.25in,clip,keepaspectratio]
{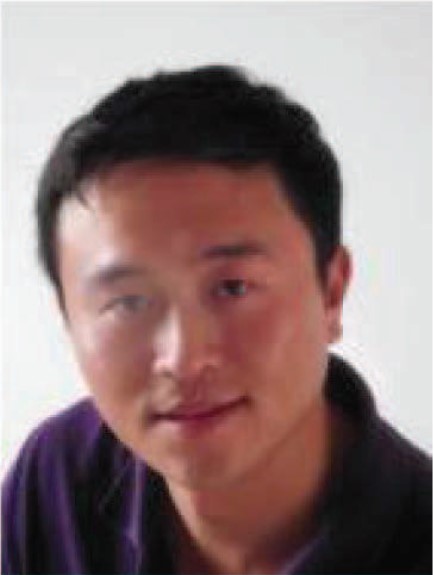}}]
{Jun Yu} (M'13) received the B.Eng. and Ph.D. degrees from Zhejiang University, Zhejiang, China.
He is currently a Professor with the School of Computer Science and Technology, Hangzhou Dianzi University, Hangzhou, China. He was an Associate Professor with the School of Information Science and Technology, Xiamen University, Xiamen, China. From 2009 to 2011, he was with Nanyang Technological University, Singapore. From 2012 to 2013, he was a Visiting Researcher at Microsoft Research Asia (MSRA). Over the past years, his research interests have included multimedia analysis, machine learning, and image processing. He has authored or coauthored more than 50 scientific articles.
Prof. Yu has (co-)chaired several special sessions, invited sessions, and workshops. He served as a program committee member or reviewer of top conferences and prestigious journals. He is a Professional Member of the Association for Computing Machinery (ACM) and the China Computer Federation (CCF).
\end{IEEEbiography}

\begin{IEEEbiography}[{\includegraphics[width=1in,height=1.25in,clip,keepaspectratio]{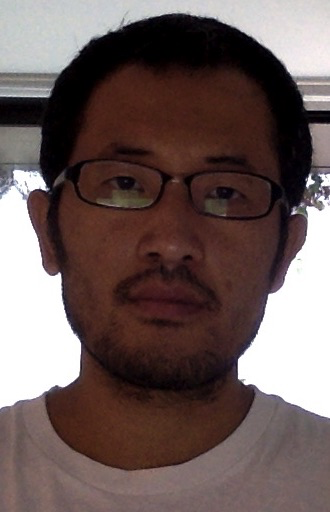}}]
{Wei Bian}
(M'14) received the B.Eng. degree in electronic engineering and the B.Sc. degree in applied mathematics in 2005, the M.Eng. degree in electronic engineering in 2007, all from the Harbin institute of Technology, harbin, China, and the PhD degree in computer science in 2012 from the University of Technology Sydney.
His research interests are pattern recognition and machine learning.
\end{IEEEbiography}

\begin{IEEEbiography}[{\includegraphics[width=1in,height=1.25in,clip,keepaspectratio]{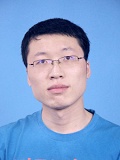}}]
{Qiang Li}
received the BEng degree in Electronics and Information Engineering, and MEng degree in Signals and Information Processing, from Huazhong University of Science and Technology (HUST), Wuhan, China, in 2010 and 2013, respectively. He is currently pursuing PhD degree at the University of Technology Sydney (UTS), Australia and a joint PhD degree at The Hong Kong Polytechnic University (PolyU), Hong Kong. His primary research involves probabilistic graphical models and its applications in computer vision, with particular interests in variational inference, structured prediction, and latent variable models.
\end{IEEEbiography}

\begin{IEEEbiography}[{\includegraphics[width=1in,height=1.25in,clip,keepaspectratio]{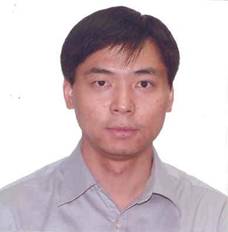}}]
{Dacheng Tao} (F¡¯15) is Professor of Computer Science and ARC Laureate Fellow in the School of Information Technologies and the Faculty of Engineering and Information Technologies, and the Inaugural Director of the UBTECH Sydney Artificial Intelligence Centre, at the University of Sydney. He mainly applies statistics and mathematics to Artificial Intelligence and Data Science. His research interests spread across computer vision, data science, image processing, machine learning, and video surveillance. His research results have expounded in one monograph and 500+ publications at prestigious journals and prominent conferences, such as IEEE T-PAMI, T-NNLS, T-IP, JMLR, IJCV, NIPS, ICML, CVPR, ICCV, ECCV, ICDM; and ACM SIGKDD, with several best paper awards, such as the best theory/algorithm paper runner up award in IEEE ICDM¡¯07, the best student paper award in IEEE ICDM¡¯13, the distinguished student paper award in the 2017 IJCAI, the 2014 ICDM 10-year highest-impact paper award, and the 2017 IEEE Signal Processing Society Best Paper Award. He received the 2015 Australian Scopus-Eureka Prize, the 2015 ACS Gold Disruptor Award and the 2015 UTS Vice-Chancellor¡¯s Medal for Exceptional Research. He is a Fellow of the IEEE, AAAS, OSA, IAPR and SPIE.
\end{IEEEbiography}

\end{document}